\documentclass[aps,Preprint,superscriptaddress]{revtex4-2}
\usepackage[utf8]{inputenc}
\usepackage{physics}
\usepackage{mathrsfs}
\usepackage[colorlinks=true,citecolor=blue,linkcolor=blue]{hyperref}
\usepackage[english]{babel}
\usepackage[T1]{fontenc}
\usepackage[left=2cm,right=2cm,top=2cm,bottom=2cm]{geometry}
\usepackage{amsmath,amssymb,amsthm,dsfont}
\usepackage{amsfonts}
\usepackage{graphicx}
\usepackage{subfig}
\usepackage{wrapfig}
\usepackage{slashed}
\usepackage{fancyhdr}
\selectfont
\usepackage{babel}

\begin{document}


\title{Relativistic elastic scattering of a muon neutrino by an electron in an elliptically polarized laser field}


\author{R. Chahri}
\affiliation{Polydisciplinary Faculty, Laboratory of Research in Physics and Engineering Sciences, Team of Modern and Applied Physics, Sultan Moulay Slimane University, Beni Mellal, 23000, Morocco.}
\author{S. El Asri}
\affiliation{Polydisciplinary Faculty, Laboratory of Research in Physics and Engineering Sciences, Team of Modern and Applied Physics, Sultan Moulay Slimane University, Beni Mellal, 23000, Morocco.}
\author{S. Mouslih}
\affiliation{ Faculté des Sciences et Techniques, Laboratoire de l’Ingénierie en Chimie et Physique de la Matière (LICPM), Beni Mellal, 23000, Morocco.}
		\affiliation{Polydisciplinary Faculty, Laboratory of Research in Physics and Engineering Sciences, Team of Modern and Applied Physics, Sultan Moulay Slimane University, Beni Mellal, 23000, Morocco.}
\author{M. Jakha}
\affiliation{Polydisciplinary Faculty, Laboratory of Research in Physics and Engineering Sciences, Team of Modern and Applied Physics, Sultan Moulay Slimane University, Beni Mellal, 23000, Morocco.}
\author{B. Manaut}
\affiliation{Polydisciplinary Faculty, Laboratory of Research in Physics and Engineering Sciences, Team of Modern and Applied Physics, Sultan Moulay Slimane University, Beni Mellal, 23000, Morocco.}
\author{S. Taj}
\email[]{s.taj@usms.ma}
\affiliation{Polydisciplinary Faculty, Laboratory of Research in Physics and Engineering Sciences, Team of Modern and Applied Physics, Sultan Moulay Slimane University, Beni Mellal, 23000, Morocco.}


\date{\today}

\begin{abstract}
Within the framework of electroweak theory, we investigate the elastic scattering process $ e^{-} + \nu_{\mu} \rightarrow e^{-} + \nu_{\mu}$ in the presence of an intense elliptically polarized laser field. We derive an analytical expression for the spin-unpolarized differential cross section using the first Born approximation and the Dirac-Volkov states to describe the incident and scattered electrons. Our results generalize those found for the linearly polarized field by Bai \textit{et al.} [Phys. Rev. A \textbf{85}, 013402 (2012)] and for the circularly polarized field by El Asri \textit{et al.} [Phys. Rev. D \textbf{104}, 113001 (2021)]. We find that the differential cross section is significantly enhanced for linear polarization and reduced for circular and elliptical polarizations.
  \end{abstract}

\maketitle
\section{Introduction}
Due to the rapid and advanced progress in the field of laser technology since the early 1960s, lasers have become widely used in industrial, medical, commercial, scientific and military domains. Lasers can provide us with sources having extreme properties in terms of energy, pulse width and wavelength, helping researchers to understand the fundamental concepts of radiation-matter interaction. Development of lasers with shorter wavelengths, shorter pulses and higher intensities continues unabated. The achievement of a maximum laser intensity of $10^{22}$ W/cm$^{2}$ \cite{yoon2019achieving} should lead to a better understanding of the behavior of various scattering processes \cite{ghatak2021lasers}. Therefore, lasers are currently indispensable tools for investigating physical processes, in particular, laser-matter interaction. The early studies of laser-assisted scattering in the nonrelativistic regime and at moderate field strengths are well established and documented in the literature. A comprehensive overview of this can be found in the books of Mittleman \cite{Mittleman}, Faisal \cite{Faisal},  Delone \cite{Delone}, Fedorov \cite{FEDOROV}  and in some recent reviews \cite{Ehlotzky2001, Francken-Joachain,ehlotzky2009fundamental}. With the advent of very powerful laser sources, it has become important to consider laser-assisted processes in the relativistic regime. Therefore, in a laser field of relativistic intensity, many processes have been studied such as Mott scattering in an elliptically and linearly polarized laser field \cite{attaourti2004mott, Li-Berakdar2003}, laser-assisted bremsstrahlung for circular and linear polarization \cite{schnez2007laser}. Electron-proton elastic scattering has been investigated in the presence of a linearly and circularly polarized laser field in \cite{Dahiri2021, wang2019relativistic, Liu2014}. In addition, in \cite{du2018new, du2018Nonlinear}, the authors studied new phenomena in laser-assisted scattering of an electron (positron) by a muon with different polarizations. There are also some papers that studied decay processes in the presence of laser field \cite{mouslihplb,MOUSLIH2020,decay1,decay2,decay3}. In this paper, using the first Born approximation, we give complete analytical and numerical results of the scattering process $e^{-}+ \nu_{\mu} \rightarrow e^{-} + \nu_{\mu}$, assisted by an elliptically polarized laser field. These results can generate those found in previous works for a circularly polarized laser field by El Asri \textit{et al.} \cite{asri2021elastic} and for a linearly polarized laser field by Bai \textit{et al.} \cite{bai2012multiphoton}. In addition, this process has been recently studied in the framework of electroweak theory for a circularly polarized laser field by \cite{asri2022}. Overall, the goal of this work is to generalize the previous research, with a more detailed calculation in the presence of an elliptically polarized laser field. We have also compared the differential cross section (DCS) in the absence and presence of a laser field at different polarizations, and examined its dependence on the laser parameters. For the elliptical polarization of the laser field, our current calculations lead to a new form of ordinary and generalized Bessel functions \cite{korsch2006two,reiss1980effect,dattoli1992generating}. The paper is organized as follows. In Sec. \ref{THEORY}, we establish the detailed analytical calculation of the S-matrix element, in the first Born approximation, of the process $e^{-}+ \nu_{\mu} \rightarrow e^{-} + \nu_{\mu}$, as well as the expression for the DCS in the presence of an external elliptically polarized laser field. Then, in Sec. \ref{NUMERICAL_RESULTS_AND_DISCUSSION}, we present numerical results for the laser-assisted DCS and discuss its dependence on the relevant parameters. Sec. \ref{CONCLUSION} is devoted to conclusions. Throughout this work, we use the natural units ($\hbar = c =1$) and the Minkowski metric tensor $g^{\mu \nu} = \text{diag}(1,-1,-1,-1)$.
\section{Theory} \label{THEORY}
\begin{figure}[hbtp]
   \centering
   \includegraphics[scale=0.3]{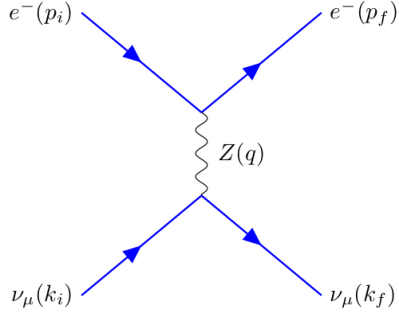}  
    \caption{The tree-level Feynman diagram at the $t$-channel of the process $ e^{-} + \nu_{\mu} \rightarrow e^{-} + \nu_{\mu} $ in the framework of electroweak theory. The time flows from left to right.}\label{Fig.Diagramme}
    \end{figure}

We consider the scattering process of a muon neutrino by an electron schematized as follows:
\begin{equation}\label{eq.1}
e^{-}(p_{i},s_{i}) + \nu_{\mu}(k_{i},t_{i}) \rightarrow e^{-}(p_{f},s_{f}) +  \nu_{\mu}(k_{f},t_{f}),
\end{equation}   
where the labels $(p_{i,f},k_{i,f})$ and $(s_{i,f},t_{i,f})$ are the associated four-momenta and spin respectively, with $i$ and $f$ stand for the initial and final states. In the framework of electroweak theory, this scattering process is mediated by the exchange of only the neutral $Z$ boson. The corresponding lowest order Feynman diagram is given by Fig.~\ref{Fig.Diagramme}.
\\
We treat this process in the presence of an elliptically polarized laser field described by the following classical four-potential \cite{attaourti2004mott}
\begin{equation}\label{eq.2}
 \begin{split}
 A^{\mu}= a_{1}^{\mu} \cos(\phi) + a_{2}^{\mu} \sin(\phi) \tan(\eta/2), 
  \end{split}
\end{equation}
where $ \phi= k^{\mu}x_{\mu} =(k.x)$, and $ \eta $ is the degree of ellipticity of the external electromagnetic (EM) field. The linear polarization is obtained for $ \eta= 0^{\circ} $, while the circular polarization is obtained for $ \eta= 90^{\circ} $. We choose the wave four-vector as $ k^{\mu}=(\omega,\textbf{k})=\omega(1,0,0,1) $. The polarization four-vectors $ a_{1}^{\mu}=(0,\textbf{a}_1)=|\textbf{a}|(0,1,0,0) $ and $ a_{2}^{\mu}=(0,\textbf{a}_2)=|\textbf{a}|(0,0,1,0)$ are along the $x$ and $y$-axis, respectively. $ a_{1}^{\mu} $ and $ a_{2}^{\mu} $ satisfy the normalization $ a_{1}^{2}=a_{2}^{2}=a^{2}=-|\textbf{a}|^{2}=-(\mathcal{E}_{0}/\omega)^{2} $ and the orthogonality conditions $ (a_{1}. a_{2})=0 $, with $ \mathcal{E}_{0} $ is the laser field strength and $ \omega $ is the frequency. Moreover, the four-potential $ A^{\mu} $ satisfies the Lorenz gauge condition $k_{\mu} A^{\mu}=0$, which implies that $ k_{\mu} a_{1}^{\mu}=k_{\mu} a_{2}^{\mu}=0 $, forcing the wave vector $ \textbf{k} $ to be along the $z$-axis. The incoming and outgoing muon neutrinos, which do not interact with the laser field, are treated as non-mass particles, described by Dirac wave functions, normalized to the volume $ V$ given by the following formula \cite{greiner1996gauge}
  \begin{equation}\label{eq.3}
 \begin{split}
 \begin{matrix} 
  \psi^{i,f}_{\nu_{\mu}}(x) = \dfrac{1}{\sqrt{2 k^{0}_{i,f} V}} u_{\nu_{\mu}}(k_{i,f},t_{i,f})~ e^{-i k_{i,f}. x}.
  \end{matrix}
 \end{split}
 \end{equation}
In the presence of a laser field, the electron obeys the following Dirac-Volkov equation \cite{volkov1935solution} 
\begin{equation}
 \begin{split}
\big\lbrace (\widehat{p_{i,f}}-e A)^{2}-m^{2}-\dfrac{1}{2} i e F_{\mu \nu} \sigma^{\mu \nu} \big\rbrace   \psi^{i,f}_{e^{-}}(x)=0,
\end{split}
 \end{equation} 
where $ e = -|e| < 0 $ is the electric charge of electron. Dressed by an elliptically polarized laser field, the incoming and outgoing electrons can be considered as Dirac-Volkov states normalized to volume $V$ \cite{volkov1935solution} 
\begin{equation}\label{eq.5}
 \begin{split}
 \psi^{i,f}_{e^{-}}(x) =\Big[ 1+ \dfrac{e}{2(k.p_{i,f})} \Big( \slashed k \slashed a_{1} \cos(\phi) + \slashed k \slashed a_{2} \sin(\phi)\tan(\eta/2) \Big) \Big]\dfrac{u_{e}(p_{i,f},s_{i,f})}{\sqrt{2 Q_{i,f} V}} e^{i S(q_{i,f},x)}, 
  \end{split}
 \end{equation}
  where 
\begin{equation}
 \begin{split} 
S(q_{i,f},x)=& -(q_{i,f}.x)-\dfrac{e(p_{i,f}.a_{1})}{(k.p_{i,f})}\sin(\phi) +\dfrac{e^{2} a^{2}}{8(k.p_{i,f})}(1-\tan^{2}(\eta/2)) \sin( 2 \phi)  +\dfrac{e(p_{i,f}.a_{2})}{(k.p_{i,f})} \tan(\eta/2)\cos(\phi),
\end{split}
 \end{equation} 
where $ u_{\nu_{\mu}}(k_{i,f},t_{i,f})$ and $u_{e}(p_{i,f},s_{i,f})$ represent the Dirac bispinors satisfying $\sum_{t_{i,f}}  \bar u_{\nu_{\mu}}(k_{i,f},t_{i,f}) u_{\nu_{\mu}}(k_{i,f},t_{i,f})=  \slashed k_{i,f} $  and  $\sum_{s_{i,f}} \bar u_{e}(p_{i,f},s_{i,f}) u_{e}(p_{i,f},s_{i,f})=  \slashed p_{i,f} + m_{e}$.  $ q_{i,f}=(Q_{i,f},\textbf{q}_{i,f})$ is the Volkov momentum of the electron in the presence of a laser field. That is
 \begin{equation}
q_{i,f}^{\mu}=p_{i,f}^{\mu} - \dfrac{e^{2} a^{2}}{4(k.p_{i,f})}(1+\tan^{2}(\eta/2)) k^{\mu}.
 \end{equation}
The square of this four-momentum shows that the mass of the dressed electron (effective mass) is proportional to the strength of the EM field as follows :
\begin{equation}
q_{i,f}^{2}= m_{*}^{2}=m_e^{2} - \dfrac{e^{2} a^{2}}{2}(1+\tan^{2}(\eta/2)),
 \end{equation}
where $m_e$ is the mass of electron, and the quantity $ m_{*} $ represents the effective mass of the electron in the elliptically polarized EM field. \\
In the first Born approximation, the transition matrix element can be expressed using Feynman rules as follows:
\begin{equation}\label{eq.7}
  \begin{split}
  S_{fi}(e^{-} \nu_{\mu} \rightarrow e^{-} \nu_{\mu}) = & -i \int d^{4}x \int d^{4}y \Big[\bar  \psi^{f}_{\nu_{\mu}}(x) \dfrac{-i g}{4 \cos(\theta_{W})} \gamma^{\mu}(1-\gamma^{5}) \psi^{i}_{\nu_{\mu}}(x) \Big] D_{\mu \nu }(x-y) \\ & \times \Big[\bar  \psi^{f}_{e}(y)\dfrac{-i g}{2 \cos(\theta_{W})} \gamma^{\nu}(g_{V}-g_{A} \gamma^{5}) \psi^{i}_{e}(y) \Big],
   \end{split}
 \end{equation}
with  $g_{V}=1/2 - 2 \sin^{2}(\theta_{W})$ and $g_{A}=-1/2$ are the coupling constants \cite{renton1990electroweak}, where $\theta_{W}$ is the Weinberg angle. Here, we choose $g_{V}=0,043$ and $g_{A}=-0,545$  \cite{greiner1996gauge}. $D_{\mu \nu }(x-y) $ is the Feynman propagator for the coupling between the $Z$-boson and the fermions given by \cite{greiner2009salam}
 \begin{equation}\label{eq.8}
  \begin{split}
  D_{\mu \nu }(x-y) = - i \int \dfrac{d^{4}q}{(2\pi)^{4}} \dfrac{e^{- i q(x-y)}}{q^{2}-M_{Z}^{2}+i M_{Z} \Gamma_{Z}} ~\Big[  g_{\mu \nu} -\dfrac{q_{\mu} q_{\nu}}{M_{Z}^{2}}   \Big],
   \end{split}
 \end{equation}
where $ M_{Z}=(91.1876 \pm 0.0021)~\text{GeV} $ is the rest mass of the $Z$-boson and $\Gamma_{Z}=(2.4952 \pm 0.0023)~\text{GeV}$ is its total decay rate \cite{particle2020review}. After inserting Eqs.~(\ref{eq.3}), (\ref{eq.5}) and (\ref{eq.8}) into Eq.~(\ref{eq.7}) and after some algebraic manipulations, we find 
	\begin{equation} \label{eq.9}
 	\begin{split}
 	S_{fi}(e^{-} \nu_{\mu} \rightarrow e^{-} \nu_{\mu})=& \dfrac{ g^{2}}{8 \cos^2(\theta_{W})~(q^{2}-M_{Z}^{2}+i M_{Z} \Gamma_{Z})}\int d^{4}xd^{4}y\dfrac{d^{4}q}{(2\pi)^{4}} \dfrac{e^{-iq(x-y)}e^{i(k_{f}-k_{i}).x}}{\sqrt{16k^{0}_{i}k^{0}_{f}Q_{i}Q_{f} V^{4}}} \\ & \times e^{i(S(q_{i},y)-S(q_{f},y))} \Big[\bar{u}_{\mu_{\mu}}(k_{f},t_{i})\gamma^{\mu}(1-\gamma_{5})u_{\nu_{\mu}}(k_{i},t_{f}) \Big] \\ & \times \Big[\bar{u}_{e}(p_{f},s_{f})\Big(1+\dfrac{e\slashed{A}\slashed{k}}{2(k.p_{f})}\Big)
 \gamma_{\mu}(g_{V}-g_{A}\gamma^{5})\Big(1+\dfrac{e\slashed{k}\slashed{A}}{2(k.p_{i})}\Big)~u_{e}(p_{i},s_{i}) \Big]. 	
 	\end{split}
 	\end{equation}
We expand the term $e^{i(S(q_{i},y)-S(q_{f},y))}$ in Eq.~(\ref{eq.9}) under the following transformation
	\begin{equation}\label{eq.10} 
 	\begin{split}
e^{i(S(q_{i},y)-S(q_{f},y))}= e^{i(q_{f}-q_{i}).y}~e^{i \big( \xi_{1} \sin(\phi) + \xi_{2} \sin(2\phi) \big)}~e^{i \xi_{3} \cos(\phi)},
 	\end{split}
 	\end{equation}
where
\begin{equation}\label{eq.11}
 	\xi_{1}= e  \Big(  \dfrac{p_{f}.a_{1}}{k.p_{f}} -\dfrac{p_{i}.a_{1}}{k.p_{i}}  \Big),~~\xi_{2}= \dfrac{-e^{2} a^{2}}{8}  \Big(  \dfrac{1}{k.p_{f}} -\dfrac{1}{k.p_{i}}  \Big) \Big( 1-\tan^{2}(\eta/2) \Big),~~\xi_{3}=-e  \Big(  \dfrac{p_{f}.a_{2}}{k.p_{f}} -\dfrac{p_{i}.a_{2}}{k.p_{i}} \Big) \tan(\eta/2).
 	\end{equation}
By introducing Eq.~(\ref{eq.10}) and using $ g^{2}/(8 \cos^2(\theta_{W}))=G_{F} M^{2}_{Z}/\sqrt{2} $, where $\textit{G}_{F}=(1.16637\pm 0.00002)\times10^{-11}~\text{MeV}^{-2}$ is the Fermi coupling constant, the expression of S-matrix becomes  
	\begin{equation} \label{eq.12}
 	\begin{split}
 	S_{fi}(e^{-} \nu_{\mu} \rightarrow e^{-} \nu_{\mu})=& \dfrac{ G_{F} M^{2}_{Z}}{\sqrt{2}~(q^{2}-M_{Z}^{2}+i M_{Z} \Gamma_{Z})}\int d^{4}xd^{4}y\dfrac{d^{4}q}{(2\pi)^{4}} \dfrac{e^{i(k_{f}-k_{i}-q).x}~e^{i(q_{f}-q_{i}+q).y}}{\sqrt{16k^{0}_{i}k^{0}_{f}Q_{i}Q_{f} V^{4}}} \\ & \times e^{i \big(\xi_{1} \sin(\phi) + \xi_{2} \sin(2\phi)\big)}   e^{i \xi_{3} \cos(\phi)}~ \Big[\bar{u}_{\mu_{\mu}}(k_{f},t_{i})\gamma^{\mu}(1-\gamma_{5})~u_{\nu_{\mu}}(k_{i},t_{f}) \Big] \\ & \times  \Big[\bar{u}_{e}(p_{f},s_{f})\Big(\chi_{0 \mu} +\chi_{1 \mu} \cos(\phi) + \chi_{2 \mu} \sin(\phi) + \chi_{3 \mu} \cos^{2}(\phi)  \Big) u_{e}(p_{i},s_{i}) \Big],
 	\end{split}
 	\end{equation}
where the quantities $\chi_{0 \mu}$, $\chi_{1 \mu}$, $\chi_{2 \mu}$ and $\chi_{3 \mu}$ are expressed as follows :
\begin{equation} \label{eq.13}
 	\begin{split}
 	&\chi_{0 \mu} = \gamma_{\mu}( g_{V}-g_{A} \gamma_{5})-2 C(p_{i}) C(p_{f})a^{2} k_{\mu} \slashed k ( g_{V}-g_{A} \gamma_{5}) \tan^{2}(\eta/2) ,\\
 	&\chi_{1 \mu} = C(p_{f}) \slashed a_{1} \slashed k \gamma_{\mu}( g_{V}-g_{A} \gamma_{5})+C(p_{i}) \gamma_{\mu}( g_{V}-g_{A} \gamma_{5}) \slashed k \slashed a_{1},\\
 	&\chi_{2 \mu} = \big[C(p_{f}) \slashed a_{2} \slashed k \gamma_{\mu}( g_{V}-g_{A} \gamma_{5})+C(p_{i}) \gamma_{\mu}( g_{V}-g_{A} \gamma_{5}) \slashed k \slashed a_{2}\big]\tan(\eta/2) ,\\
 	&\chi_{3 \mu}= -2 C(p_{i}) C(p_{f})a^{2} k_{\mu} \slashed k ( g_{V}-g_{A} \gamma_{5}) \big(1- \tan^{2}(\eta/2) \big),
 	\end{split}
 	\end{equation}
where $ C(p_{i,f}) = e/[2(k. p_{i,f})] $. Now, we use a transformation, known as the ordinary and generalized Jacobi-Anger identity, involving ordinary $J_{s}(\xi_{3})$ and generalized $ B_{s}(\xi_{1},\xi_{2})$ Bessel functions \cite{dattoli1992generating}:
\begin{equation}
 	\begin{split}
  e^{i(\xi_{1} \sin(\phi) + \xi_{2} \sin(2\phi))}&=\sum_{s=-\infty}^{+\infty} B_{s}(\xi_{1},\xi_{2})~e^{i s \phi} ,\\
  e^{i \xi_{3} \cos(\phi)}&=\sum_{s=-\infty}^{+\infty} i^{s} j_{s}(\xi_{3})~e^{i s \phi},\\
	\end{split}
 	\end{equation}
where $s$, the order of Bessel functions, is commonly interpreted as the number of photons exchanged between the two particles involved in our scattering process and the laser field. Explicitly, we apply the following transformation 
\begin{equation}
 \begin{split}
\begin{Bmatrix}
1\\
\cos(\phi)\\
\sin(\phi) \\
\cos^{2}(\phi)
\end{Bmatrix} 
  e^{i(\xi_{1} \sin(\phi) + \xi_{2} \sin(2\phi))}~e^{i \xi_{3} \cos(\phi)}  =\sum_{s=-\infty}^{+\infty} \begin{Bmatrix}
B_{s}(\xi_{3})\\
B_{1s}(\xi_{3})\\
B_{2s}(\xi_{3})\\
B_{3s}(\xi_{3})
\end{Bmatrix}
B_{s}(\xi_{1},\xi_{2})~e^{i 2s \phi},
 \end{split}
 \end{equation} 
where the coefficients $B_{s}(\xi_{3})$, $B_{1s}(\xi_{3})$, $B_{2s}(\xi_{3})$, $B_{3s}(\xi_{3})$ and $ B_{s}(\xi_{1},\xi_{2}) $ are expressed in terms of ordinary Bessel functions as follows \cite{lotstedt2009recursive,dattoli1992generating}: 
    \begin{equation}
 	\begin{split}
 	B_{s}(\xi_{3}) &= i^{s} j_{s}(\xi_{3}),\\
 	B_{1s}(\xi_{3}) &=\big( i^{s-1} j_{s-1}(\xi_{3})+ i^{s+1} j_{s+1}(\xi_{3})\big)/2 ,\\
 	B_{2s}(\xi_{3}) &=\big( i^{s-1} j_{s-1}(\xi_{3})- i^{s+1} j_{s+1}(\xi_{3})\big)/2 i ,\\
 	B_{3s}(\xi_{3})&=\big(2 i^{s} j_{s}(\xi_{3})+ i^{s-2} j_{s-2}(\xi_{3})+ i^{s+2} j_{s+2}(\xi_{3})\big)/4 ,\\
 	B_{s}(\xi_{1},\xi_{2})&= \sum_{\lambda=-\infty}^{+\infty} j_{s-2\lambda}(\xi_{1}) j_{\lambda}(\xi_{2}) .
 	\end{split}
 	\end{equation}
After integration over space-time and $d^{4}q$, and after some algebraic manipulations, we can decompose the transition matrix element into a series of terms in the form of ordinary and generalized Bessel functions
	\begin{equation} \label{eq.17}
 	\begin{split}
 	S_{fi}(e^{-} \nu_{\mu} \rightarrow e^{-} \nu_{\mu})= \sum_{s=-\infty}^{+\infty}  \dfrac{G_{F} M_{Z}^{2}}{\sqrt{2}(q^{2}-M_{Z}^{2}+i M_{Z} \Gamma_{Z})}  \dfrac{(2\pi)^{4}~\delta^{4}(q_{f}-q_{i}+k_{f}-k_{i}+2 s k) }{\sqrt{16k^{0}_{i}k^{0}_{f}Q_{i}Q_{f} V^{4}}} 
~M^{(s)}_{f i},
 	\end{split}
 	\end{equation} 	
where $q = q_{i}-q_{f}-2 s k$ is the relativistic four-momentum transfer in the presence of the EM field. The quantity $ M^{(s)}_{f i} $ in Eq.~(\ref{eq.17}) is defined by 
\begin{equation} 
 	\begin{split}
M^{(s)}_{f i}= \big[\bar{u}_{\nu_{\mu}}(k_{f},t_{i})\gamma^{\mu}(1-\gamma_{5})u_{\nu_{\mu}}(k_{i},t_{f}) \big] \big[\bar{u}_{e}(p_{f},s_{f}) \Gamma_{\mu}^{(s)} u_{e}(p_{i},s_{i}) \big],
\end{split}
 	\end{equation}
where 
\begin{equation} 
 	\begin{split}
\Gamma_{\mu}^{(s)}=\big[\chi_{0 \mu}  B_{s}(\xi_{3})+ \chi_{1 \mu} B_{1s}(\xi_{3})+ \chi_{2 \mu} B_{2s}(\xi_{3}) +\chi_{3 \mu} B_{3s}(\xi_{3}) \big]B_{s}(\xi_{1},\xi_{2}).
\end{split}
 	\end{equation}
The DCS can be obtained by summing over the final spin states and averaging over the initial ones. Note that electrons can be in two spin states, while neutrinos are only in one negative helicity state \cite{greiner1996gauge}. Therefore, we divide the square of the matrix element $ |S_{fi}|^{2}$ by the incident particle flux $ |J_{inc}| $ and the observation time interval $ T $, and multiply it by the density of the final states. The unpolarized DCS can then be written as follows:
\begin{equation}
d\bar{\sigma}(e^{-} \nu_{\mu} \rightarrow e^{-} \nu_{\mu})=~ \int V \dfrac{d^{3}q_{f}}{(2\pi)^{3}}~ \int V \dfrac{d^{3}k_{f}}{(2\pi)^{3}}~\dfrac{1}{2} \sum_{t_{i,f},s_{i,f}}\dfrac{|S_{fi}(e^{-} \nu_{\mu} \rightarrow e^{-} \nu_{\mu})|^{2}}{ T |J_{inc}|},
\end{equation}
where $|J_{inc}|=(k_{i}.q_{i})/(k^{0}_{i} Q_{i} V )$ denotes the current of the incoming particles in the laboratory system. Applying the following relations: $d^{3}q_{f}=|\textbf{q}_{f}|^{2}d|\textbf{q}_{f}|d\Omega $ and $\delta^{4}(q_{f}-q_{i}+k_{f}-k_{i}+ 2 s k)= \delta^{0}(Q_{f}+k^{0}_{f}-Q_{i}-k^{0}_{i} + 2 s \omega)\delta^{3}(\textbf{q}_{f}+\textbf{k}_{f}-\textbf{q}_{i}-\textbf{k}_{i}+2 s \textbf{k})$ and taking simplifications, we obtain 
\begin{equation}
\begin{split}
\dfrac{d\bar{\sigma}}{d\Omega}(e^{-} \nu_{\mu} \rightarrow e^{-} \nu_{\mu}) = & \sum_{s=-\infty}^{+\infty} \dfrac{\textit{G}_{F}^{2}~M_{Z}^{4}}{64(2\pi)^{2}((q^{2}-M_{Z}^{2})^{2}+ M_{Z}^{2} \Gamma_{Z}^{2})(k_{i}.q_{i})}  \int\dfrac{|\textbf{q}_{f}|^{2}d|\textbf{q}_{f}|}{k^{0}_{f}Q_{f}} \delta^{0}(Q_{f}+k^{0}_{f}-Q_{i}-k^{0}_{i} + 2 s \omega)\\ 
& \times \sum_{s'_{i,f},s_{i,f}}|M_{fi}^{(s)}|^{2}\bigg|_{\textbf{q}_{f}+\textbf{k}_{f}-\textbf{q}_{i}-\textbf{k}_{i}+2 s \textbf{k}=0}.
\end{split}
\end{equation}
Using the following relation \cite{greiner2009salam}
\begin{equation}
 \begin{split}
 \int dx f(x) \delta(g(x))=\dfrac{f(x)}{|g'(x)|}\Big|_{g(x)=0},
 \end{split}
 \end{equation}
we can perform the remaining integral over $d|\textbf{q}_{f}|$. Therefore, the summed differential cross section (SDCS) can be decomposed into a series of discrete individual differential cross section (IDCS) for different numbers of photons exchanged. This yields
\begin{equation}\label{eq.23}
 \begin{split}
\Big(\dfrac{ d \bar \sigma}{d\Omega}\Big)^{\text{with laser}}(e^{-} \nu_{\mu} \rightarrow e^{-} \nu_{\mu})= \sum_{s=-\infty}^{+\infty}  \dfrac{ d\sigma^{(s)}}{d\Omega},
 \end{split}
\end{equation}
where the IDCS can be written as
\begin{equation} \label{eq.24}
 \begin{split}
 \dfrac{ d\sigma^{(s)}}{d\Omega } =  \dfrac{\textit{G}_{F}^{2}~M_{Z}^{4}}{64(2\pi)^{2}((q^{2}-M_{Z}^{2})^{2}+ M_{Z}^{2} \Gamma_{Z}^{2})(k_{i}.q_{i}) k^{0}_{f} Q_{f}}  \dfrac{|\textbf{q}_{f}|^{2} }{\Big|g'(|\textbf{q}_{f}|)\Big|} \times    \sum_{s_{i},s_{f}} \sum_{t_{i},t_{f}} |M_{f i}^{(s)}|^{2}\bigg|_{\textbf{q}_{f}+\textbf{k}_{f}-\textbf{q}_{i}-\textbf{k}_{i}+2 s \textbf{k}=0},
 \end{split}
\end{equation}
where
\begin{equation}
\begin{split}
g'(|\textbf{q}_{f}|)= \dfrac{|\textbf{q}_{f}|}{Q_{f}} + \dfrac{|\textbf{q}_{f}|+ k^{0}_{i} \cos(\theta_{f})-|\textbf{q}_{i}| F(\theta_{i},\theta_{f},\varphi_{i},\varphi_{f}) +2 s \omega \cos(\theta_{f}) }{k^{0}_{f}},
\end{split}
\end{equation}
with
\begin{equation}
\begin{split}
 F(\theta_{i},\theta_{f},\varphi_{i},\varphi_{f})= \sin(\theta_{i}) \cos(\varphi_{i}) \sin(\theta_{f}) \cos(\varphi_{f})+ \sin(\theta_{i}) \sin(\varphi_{i}) \sin(\theta_{f}) \sin(\varphi_{f})+ \cos(\theta_{i})\cos(\theta_{f}).
\end{split}
\end{equation}
The sum over spin can be converted to trace calculation as follows:
\begin{equation} \label{eq.27}
\begin{split}
  \sum_{s_{i},s_{f}} \sum_{t_{i},t_{f}} |M_{f i}^{(s)}|^{2} = \text{Tr}\big[\slashed k_{f}  \gamma^{\mu}(1-\gamma_{5})\slashed k_{i}  \gamma^{\nu}(1-\gamma_{5}) \big]\text{Tr}\big[(\slashed p_{f}+ m_{e}) \Gamma_{\mu}^{(s)}  (\slashed p_{i} + m_{e}) \bar \Gamma_{\nu}^{(s)} \big],
  \end{split} 
 \end{equation}
where
\begin{equation} 
 	\begin{split}
 & \bar \Gamma_{\nu}^{(s)} = \gamma^{0}  \Gamma_{\nu}^{(s) \dagger} \gamma^{0}, 	\\
 & \bar \Gamma_{\nu}^{(s)}=\big[\bar \chi_{0 \nu}  B^{*}_{s}(\xi_{3})+\bar \chi_{1 \nu} B^{*}_{1s}(\xi_{3})+ \bar \chi_{2 \nu} B^{*}_{2s}(\xi_{3}) +\bar \chi_{3 \nu} B^{*}_{3s}(\xi_{3}) \big]~B^{*}_{s}(\xi_{1},\xi_{2}),
\end{split}
 	\end{equation}
and
\begin{equation}
 	\begin{split}
 	&\bar \chi_{0 \nu} = \gamma_{\nu}( g_{V}-g_{A} \gamma_{5})-2 C(p_{i}) C(p_{f})a^{2} k_{\nu} \slashed k ( g_{V}-g_{A} \gamma_{5}) \tan^{2}(\eta/2) ,\\
 	&\bar \chi_{1 \nu} = C(p_{f}) \gamma_{\nu}( g_{V}-g_{A} \gamma_{5}) \slashed k \slashed a_{1} +C(p_{i}) \slashed a_{1} \slashed k \gamma_{\nu}( g_{V}-g_{A} \gamma_{5}),\\
 	&\bar \chi_{2 \nu} =\big[ C(p_{f}) \gamma_{\nu}( g_{V}-g_{A} \gamma_{5}) \slashed k \slashed a_{2}+ C(p_{i}) \slashed a_{2} \slashed k \gamma_{\nu}( g_{V}-g_{A} \gamma_{5})\big]\tan(\eta/2) ,\\
 	&\bar \chi_{3 \nu}= -2 C(p_{i}) C(p_{f})a^{2} k_{\nu} \slashed k ( g_{V}-g_{A} \gamma_{5}) \big(1- \tan^{2}(\eta/2) \big).
 	\end{split}
 	\end{equation}
The FeynCalc program \cite{feyncalc} is used to compute the traces in Eq.(\ref{eq.27}). The result obtained is written in terms of the coefficients $\Lambda_{1}$ to $\Lambda_{16}$ as follows: 
\begin{equation}\label{trace-result}
\begin{split}
\sum_{s_{i},s_{f}} \sum_{t_{i},t_{f}} |M_{f i}^{(s)}|^{2}=& \big[ \Lambda_{1} |B_{s}(\xi_{3})|^{2} + \Lambda_{2} |B_{1s}(\xi_{3})|^{2} +\Lambda_{3} |B_{2s}(\xi_{3})|^{2} +  \Lambda_{4}~ |B_{3s}(\xi_{3})|^{2} + \Lambda_{5} B_{s}(\xi_{3}) B^{*}_{1s}(\xi_{3})\\ & + \Lambda_{6}B_{1s}(\xi_{3}) B^{*}_{s}(\xi_{3})+ \Lambda_{7} B_{s}(\xi_{3}) B^{*}_{2s}(\xi_{3}) + \Lambda_{8}~ B_{2s}(\xi_{3}) B^{*}_{s}(\xi_{3})  + \Lambda_{9} B_{s}(\xi_{3}) B^{*}_{3s}(\xi_{3})
\\ &  + \Lambda_{10}~ B_{3s}(\xi_{3}) B^{*}_{s}(\xi_{3})  + \Lambda_{11}~ B_{1s}(\xi_{3}) B^{*}_{2s}(\xi_{3}) + \Lambda_{12}~ B_{2s}(\xi_{3}) B^{*}_{1s}(\xi_{3})  + \Lambda_{13} B_{1s}(\xi_{3}) B^{*}_{3s}(\xi_{3})
\\ & + \Lambda_{14}~ B_{3s}(\xi_{3}) B^{*}_{1s}(\xi_{3}) + \Lambda_{15}~ B_{2s}(\xi_{3}) B^{*}_{3s}(\xi_{3}) + \Lambda_{16} B_{3s}(\xi_{3}) B^{*}_{2s}(\xi_{3})     \big] |B_{s}(\xi_{1},\xi_{2})|^{2} . 
  \end{split}
\end{equation}
We give the explicit expression of first four coefficients in the Appendix.
\section{Numericla results and discussion} \label{NUMERICAL_RESULTS_AND_DISCUSSION}
\begin{figure}[hbtp]
   \centering
   \includegraphics[scale=0.3]{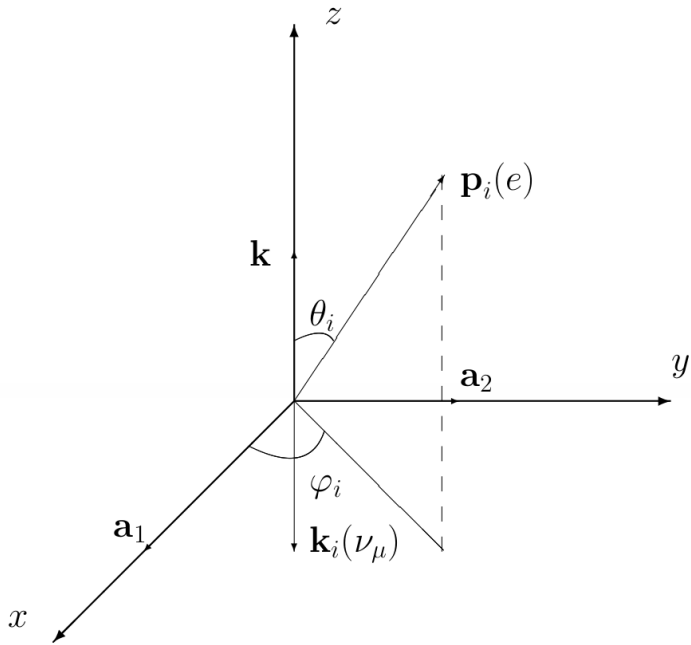}  
    \caption{Geometry of the coordinate system for the scattering process. The incident electron $e^{-}(p_{i})$ moves in a spherical coordinate geometry ($\theta_{i}$, $\varphi_{i}$), while the muon neutrino $\nu_{\mu}(k_{i})$ remains in the direction opposite to the $z$-axis. The outgoing muon neutrino and final electron move in space ($xyz$).}\label{fig.geometric}
\end{figure}

\begin{figure}[hbtp]
\centering
   \includegraphics[scale=0.45]{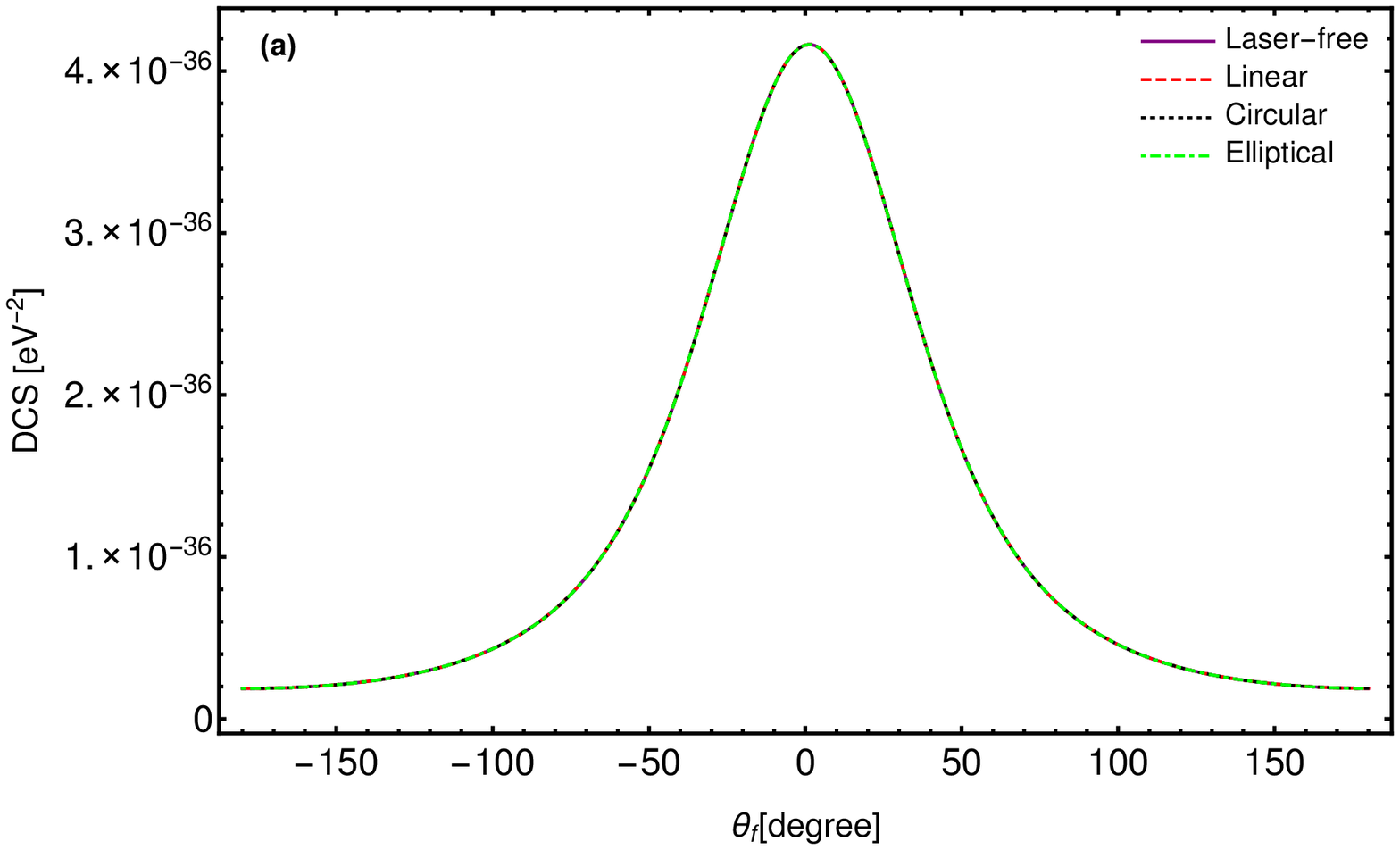}  
   \includegraphics[scale=0.45]{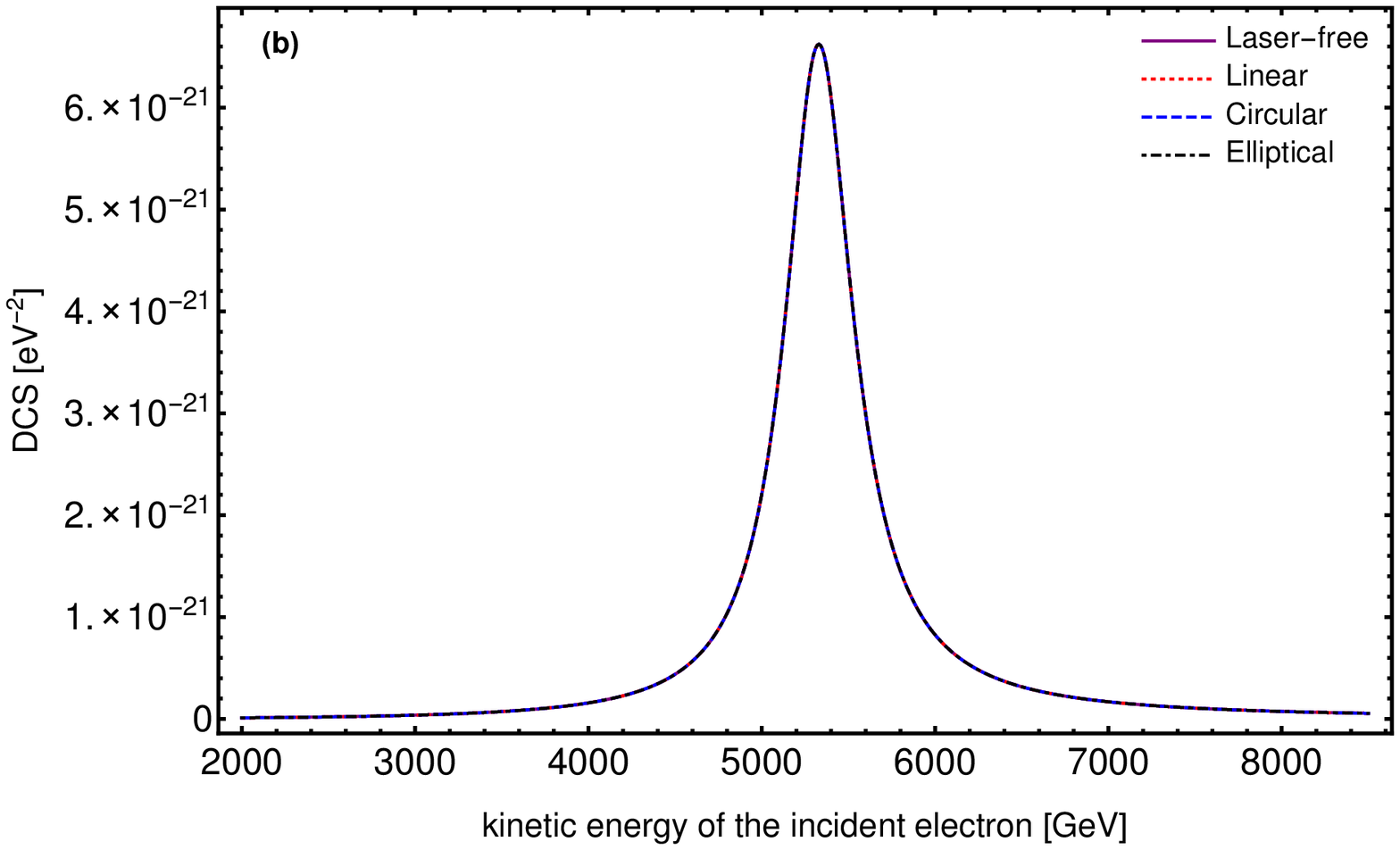}  
  \caption{The two DCSs with and without laser, illustrated as a function of the final scattering angle $ \theta_{f}$ and the kinetic energy of the incident electron for an electric field strength $ \mathcal{E}_{0}=0~\text{V/cm}$ and without any exchange of photons $(s=0)$. The remaining parameters are chosen as follows:  (a) $E_{\nu}^{\text{kin}}=0.5\times10^{-3}$ GeV, $E_{e}^{\text{kin}}=10^{-3}$ GeV,  $\theta_{i}=1^{\circ},~\varphi_{i}=\varphi_{f}=0^{\circ} $; (b) $E_{\nu}^{\text{kin}}=10$ GeV, $\theta_{i}=\varphi_{i}=\varphi_{f}=1^{\circ}$, $\theta_{f}=2^{\circ}$.}\label{fig.DCS}
    \end{figure}

\begin{figure}[hbtp]
 \centering
\includegraphics[scale=0.55]{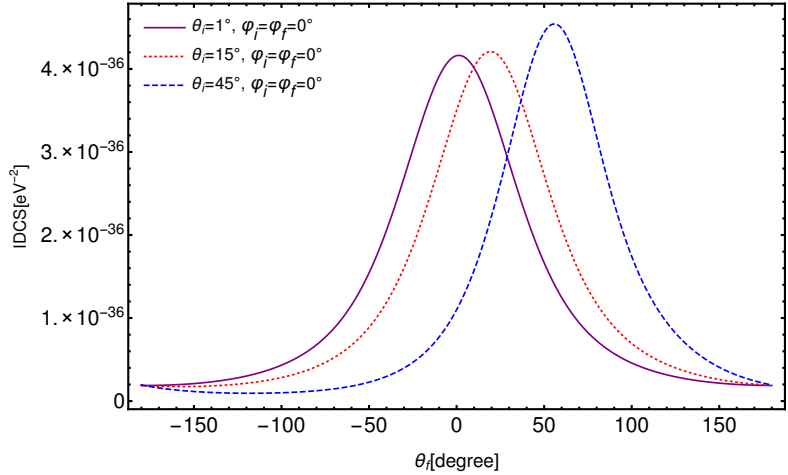}
\caption{The variations of the IDCS as a function of scattering angle $\theta_{f}$ for differents angles of the incident electron. The laser field strength and the number of photons are, respectively $\mathcal{E}_{0}=0$ V/cm and $s=0$. The kinetic energy of the incident muon neutrino and the incoming electron are, respectively, $E_{\nu}^{\text{kin}}=0.5\times10^{-3}$ GeV and $E_{e}^{\text{kin}}=10^{-3}$ GeV.}\label{fig.DCS-4}
\end{figure}

\begin{figure}[hbtp]
\centering
   \includegraphics[scale=0.5]{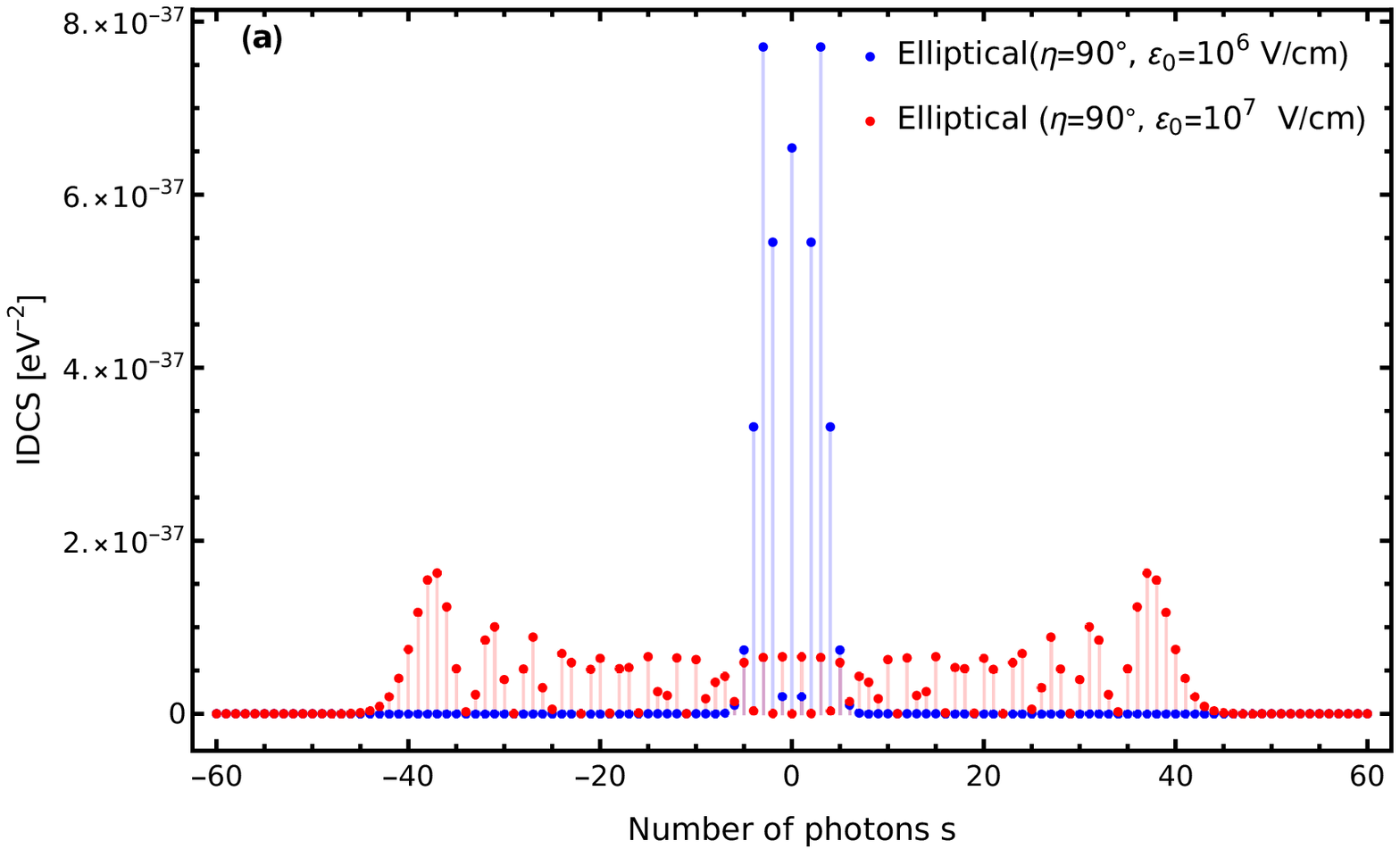}  
 \includegraphics[scale=0.55]{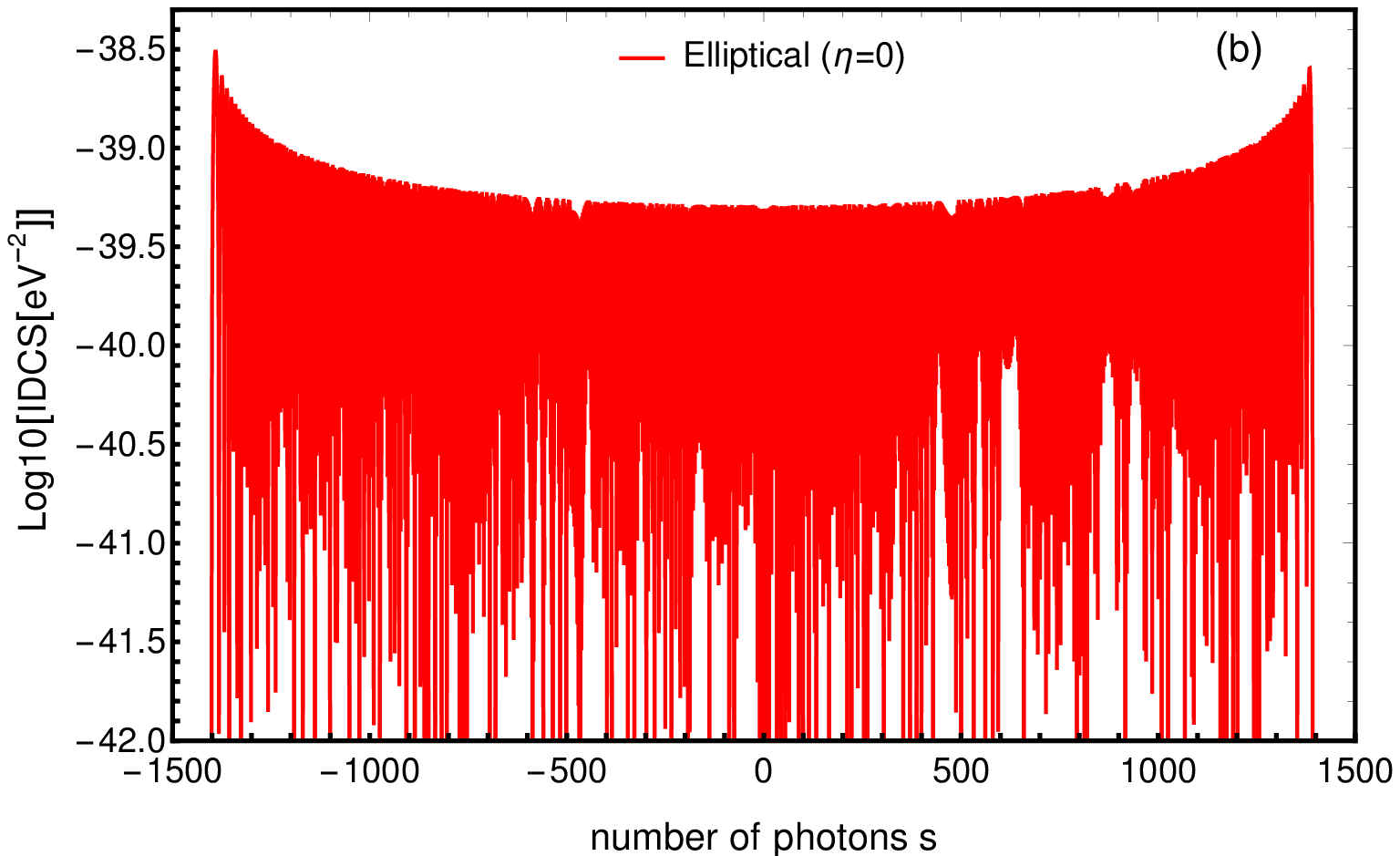}  
   \caption{The IDCS as a function of the number of photons exchanged. The free parameters are (\textbf{a}) $E_{\nu}^{\text{kin}}=0.5\times10^{-3}$ GeV, $E_{e}^{\text{kin}}=10^{-3}$ GeV, $\hbar\omega=1.17~\text{eV}$, $\theta_{i}=1^{\circ}$, $\theta_{f}=\varphi_{i}=\varphi_{f}=0^{\circ}$ (to be compared with \cite{asri2021elastic}); and (\textbf{b}) $E_{\nu}^{\text{kin}}=0.5\times10^{-3}$ GeV, $E_{e}^{\text{kin}}=10^{-3}$ GeV, $\mathcal{E}_{0}=5.18\times10^{7}~\text{V/cm}$, $\hbar\omega=1.17~\text{eV}$, $\theta_{i}=45^{\circ}$, $\varphi_{i}=\varphi_{f}=0^{\circ}$, $\theta_{f}=1^{\circ}$ (to be compared with \cite{bai2012multiphoton}).}\label{fig.IDCS-comparaison}
    \end{figure}

\begin{figure}[hbtp]
 \centering
\includegraphics[scale=0.45]{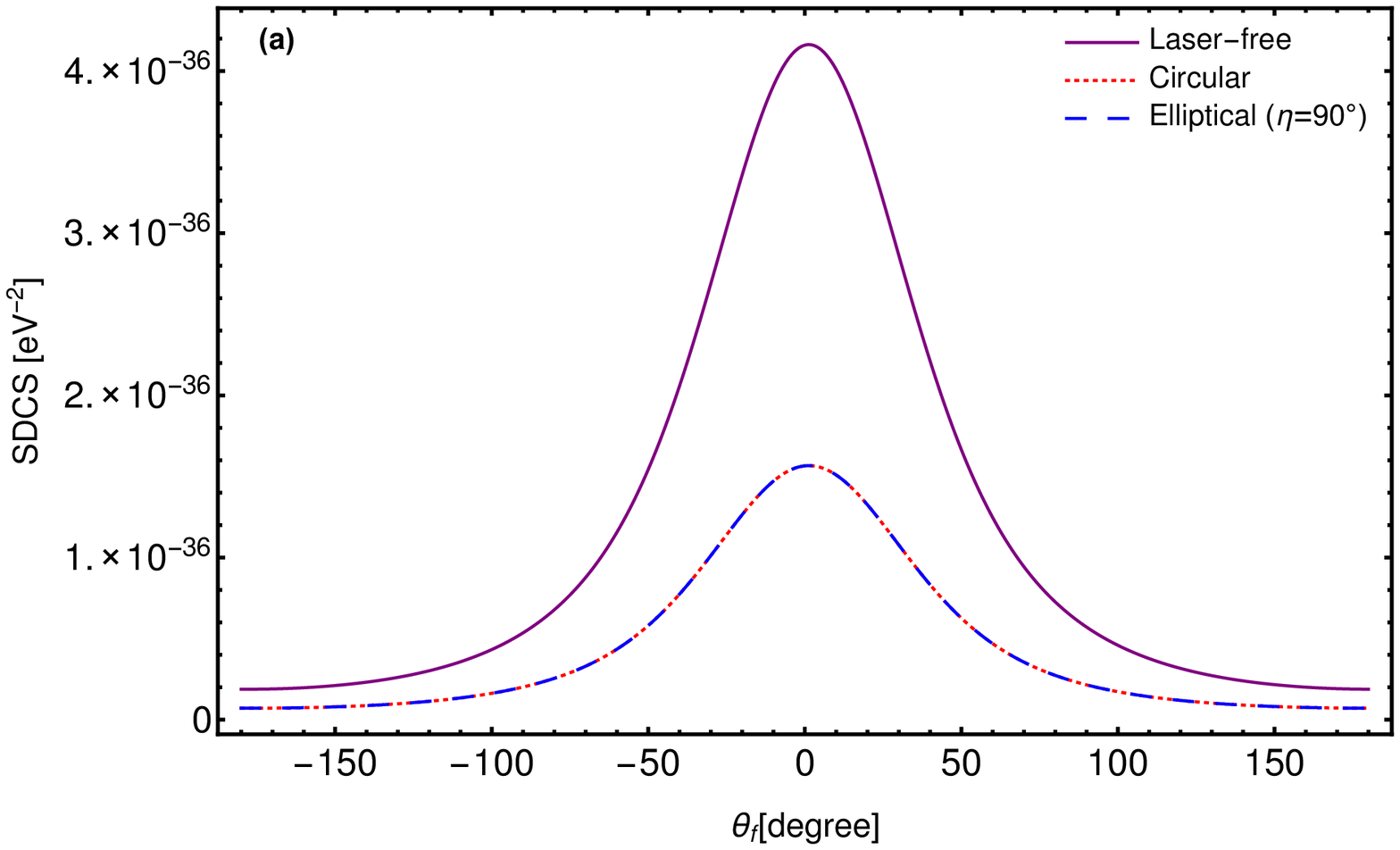}
\includegraphics[scale=0.45]{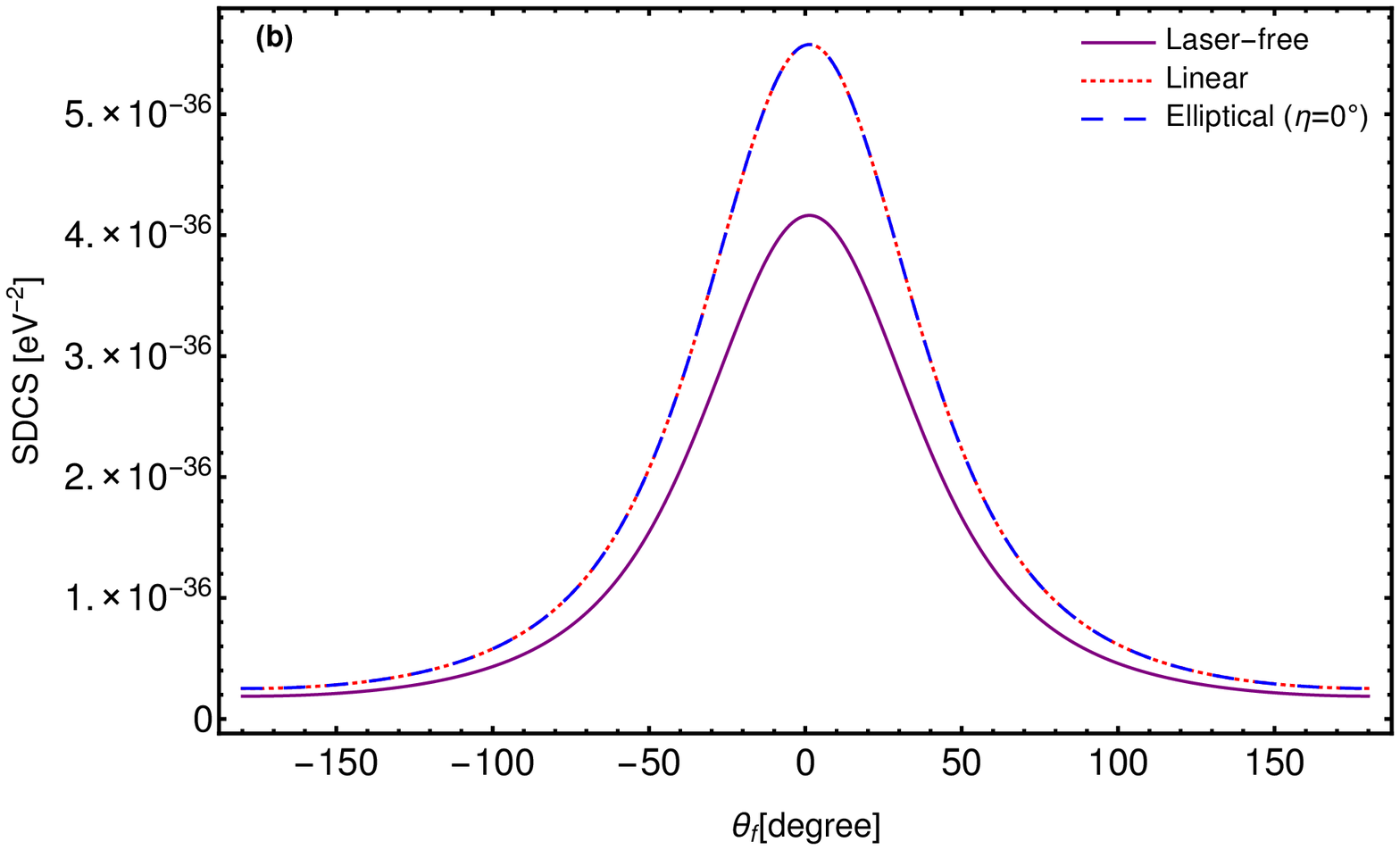}
\caption{Dependence of SDCS on the scattering angle $\theta_{f}$ for various polarizations of the EM field, taking the laser parameters such as: $\mathcal{E}_{0}=10^{8}~\text{V/cm}$, $\hbar\omega=1.17 \text{eV}$. The parameters of the geometry are taken as $\theta_{i}=\varphi_{i}=\varphi_{f}=1^{\circ}$ and the remaining parameters are selected as shown in Fig.~\ref{fig.IDCS-comparaison}.}\label{fig.SDCS-comparaison-6}
\end{figure}

In this section, we will present the numerical results obtained and discuss their physical interpretation. We should focus on experimentally measurable quantities, in particular the behavior of the DCS assisted by an external elliptically polarized EM field. We set the incident electron momentum $\textbf{p}_{i}$ and outgoing $\textbf{p}_{f}$ in a general geometry with spherical coordinates $\theta_{i}$, $\theta_{f}$, $\varphi_{i}$ and $\varphi_{f}$. Except for Figs.~\ref{fig.DCS}, \ref{fig.DCS-4} and \ref{fig.IDCS-comparaison}, we have chosen $\theta_{i}=\varphi_{i}=\varphi_{f}=1^{\circ}$ in all the results obtained. For the incident muon neutrino, the momentum $\textbf{k}_{i}$ remains in the direction opposite to the $z$ axis. Except for Fig.~\ref{fig.DCS}, we fix the kinetic energy of the incident electron at $E_{e}^{\text{kin}}=10^{-3}$ GeV, and the initial energy of the muon neutrino is chosen as the mass of an electron at rest, $E_{\nu}^{\text{kin}}=0.5\times10^{-3}$ GeV. The direction of the field wave vector $ \textbf{k}$ is along the $z$-axis, while the polarization vectors $\textbf{a}_{1}$ and $\textbf{a}_{2}$ perpendicular to $ \textbf{k}$ are along the $x$ and $y$-axes, respectively, as illustrated in Fig.~\ref{fig.geometric}. Every $d\sigma^{(s)}/ d\Omega$ given by Eq.~(\ref{eq.24}), considering the four-momentum conservation, can be interpreted as the IDCS that describes the scattering process for each number of photons $s$ ($ s > 0 $ for absorption and $ s < 0 $ for emission). Summing over a number of exchanged photons $s$, we obtain the SDCS given by Eq.~(\ref{eq.23}).    
We will show how the DCS, in the presence and absence of the laser field, varies as a function of the final scattering angle $\theta_{f}$, the kinetic energy of the incident electron $E^{kin}_{e}$ and various incident angles of the electron $\theta_{i}$. Then, we make comparisons of the IDCS and SDCS with other research papers. Afterward, we show how the IDCS depends on the number of photons exchanged $s$, on the geometry, on the parameters characterizing the EM field ($\mathcal{E}_{0}$, $\omega$) and on the kinetic energy of the incident electron. Before finishing, we illustrate the variation of the SDCS as a function of the final scattering angle $\theta_{f}$ and the kinetic energy of the incident electron $E^{kin}_{e}$ at different polarizations, frequencies, and electric field strengths $ \mathcal{E}_{0}$. Finally, we show how it evolves as a function of the electric field strength at different polarizations.
We start our discussion with something we are used to do in such processes occurring in an external EM field. That is we make sure that the DCS in the presence of the laser field with different polarizations is exactly equal to that in the absence of the laser field when the laser parameters tend to zero. In Fig.~\ref{fig.DCS}, we illustrate the comparison between the laser-assisted DCS of the electron-muon neutrino scattering in the framework of electroweak theory,  given by Eq.(\ref{eq.24}), and the corresponding one in the absence of the laser field for a geometry $\theta_{i}=1^{\circ}$, $\varphi_{i}=\varphi_{f}=0^{\circ}$ in Fig.~\ref{fig.DCS}(a), and $\theta_{i}=\varphi_{i}=\varphi_{f}=1^{\circ}$, $\theta_{f}=2^{\circ}$ in Fig.~\ref{fig.DCS}(b).
The comparison allows us to verify our results by taking the limit of the electric field strength $ \mathcal{E}_{0}=0~\text{V/cm}$ and of the number of photons exchanged $(s=0)$, where they all tend to zero. Regarding the laser field strength and frequency, these are parameters that characterize the external EM field, while the number of photons exchanged $s$ appeared due to the introduction of ordinary and generalized Bessel functions in our theoretical calculation. We note that the four graphs shown in Figs.~\ref{fig.DCS}(a) and \ref{fig.DCS}(b) are so identical as they are indistinguishable for all final scattering angles $\theta_{f}$ and kinetic energies $E^{kin}_{e}$ of the incident electron. This proves the consistency and validity of our theoretical calculations. In the next step, we will try to see the effect of the geometry in Fig.~(\ref{fig.DCS-4}), which displays the variations of the DCS in the absence of the laser field as a function of the final scattering angle $\theta_{f}$. We observe that the graphs of the DCS are distinct, which clearly shows that the geometry influences the angular distribution of the DCS. We also observe that as the incidence angle $\theta_{i}$ increases, the pic of the DCS increases with a shift towards large final scattering angles. 
Another important remark is that the electron has a high probability to be scattered under a final angle $\theta_{f}$ with a value approximately approaching the incidence angle $(\theta_{f}\approx \theta_i)$. For example, if the incidence angle is $\theta_{i}=15^{\circ}$, the peak is located around the final scattering angle $\theta_{f}\approx 15^{\circ}$. In the same context, and to highlight the correctness and accuracy of our calculations in the presence of an EM field with elliptical polarization, this calculation which is general allows us to find the results previously obtained in two research papers that deal with the same scattering process in the presence of a laser field with circular \cite{asri2021elastic} or linear \cite{bai2012multiphoton} polarization. We begin with Fig.~\ref{fig.IDCS-comparaison}(a) which presents the variations of the IDCS as a function of the number of photons exchanged $s$ with a degree of ellipticity $\eta=90$ (circular polarization) for two different field strengths $\mathcal{E}_{0}$ and for a frequency $\hbar\omega=1.17~\text{eV}$. This is the same envelope obtained in a previous paper by El Asri \textit{et al.} (see Fig.~2(a) in \cite{asri2021elastic}). Thus, the theoretical formalism adopted here is general and can lead to all the results obtained in \cite{asri2021elastic}. For linear polarization, we display, in Fig.~\ref{fig.IDCS-comparaison}(b), the IDCS as a function of the number of photons exchanged $s$ with a degree of ellipticity $\eta=0$ for the laser field strength $\mathcal{E}_{0}=5.18\times10^{7}~\text{V/cm}$ and frequency $\hbar\omega=1.17~\text{eV}$.
We observe, that a large number of photons are exchanged between the laser field and the scattering process, and the cutoff number is about $s=\pm 1400$. Comparing, this figure with the one obtained by Bai \textit{et al.} (see Fig.~2(a) in \cite{bai2012multiphoton}), we get the same result.  
Another comparison concerns Fig.~\ref{fig.SDCS-comparaison-6}, which displays the dependence of the SDCS given in Eq.~(\ref{eq.23}) on the final scattering angle $\theta_{f}$ for different polarizations of the EM field. In Fig.~\ref{fig.SDCS-comparaison-6}(a), we show the variation of the SDCS as a function of the final scattering angle $\theta_{f}$ for the degree of ellipticity $\eta=90$.
We obtain a symmetrical graph, which presents a peak in the vicinity of $\theta_{f}=0$. Furthermore, the SDCS remains lower than the DCS without a laser.
In Fig.~\ref{fig.SDCS-comparaison-6}(b), we represent the same graphs, but with a degree of ellipticity $\eta=0$. In this case, we see an enhancement of the SDCS compared to the DCS without laser, which is consistent with previous research in the case of linear polarization \cite{du2018new,du2018Nonlinear,schnez2007laser}.
After a detailed discussion and a comparison of the results obtained with the previous work, let us see what happens if we introduce an external EM field with elliptical polarization having a degree of ellipticity $\eta=120^{\circ} $.
\begin{figure}[hbtp]
\centering
   \includegraphics[scale=0.45]{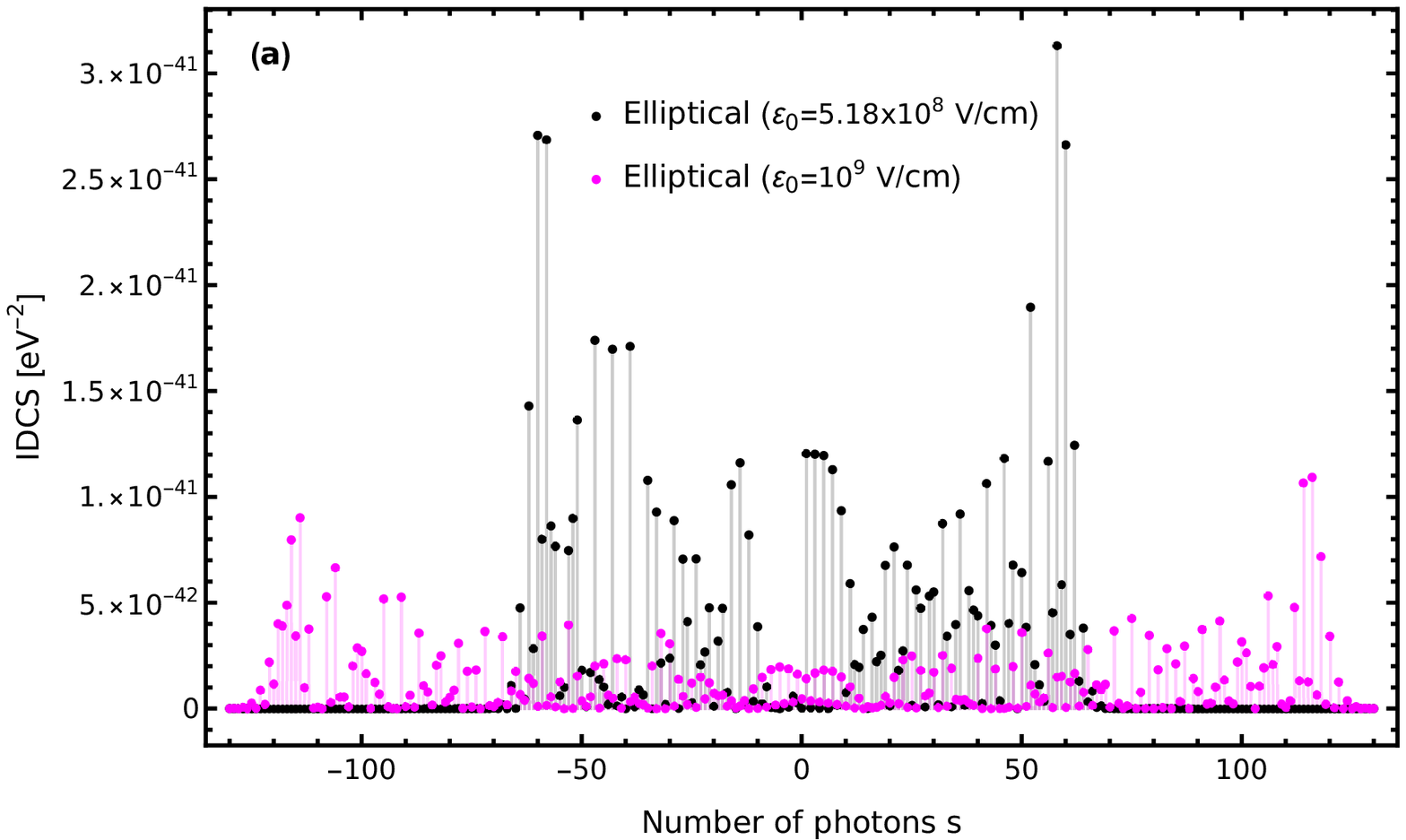}  
   \includegraphics[scale=0.45]{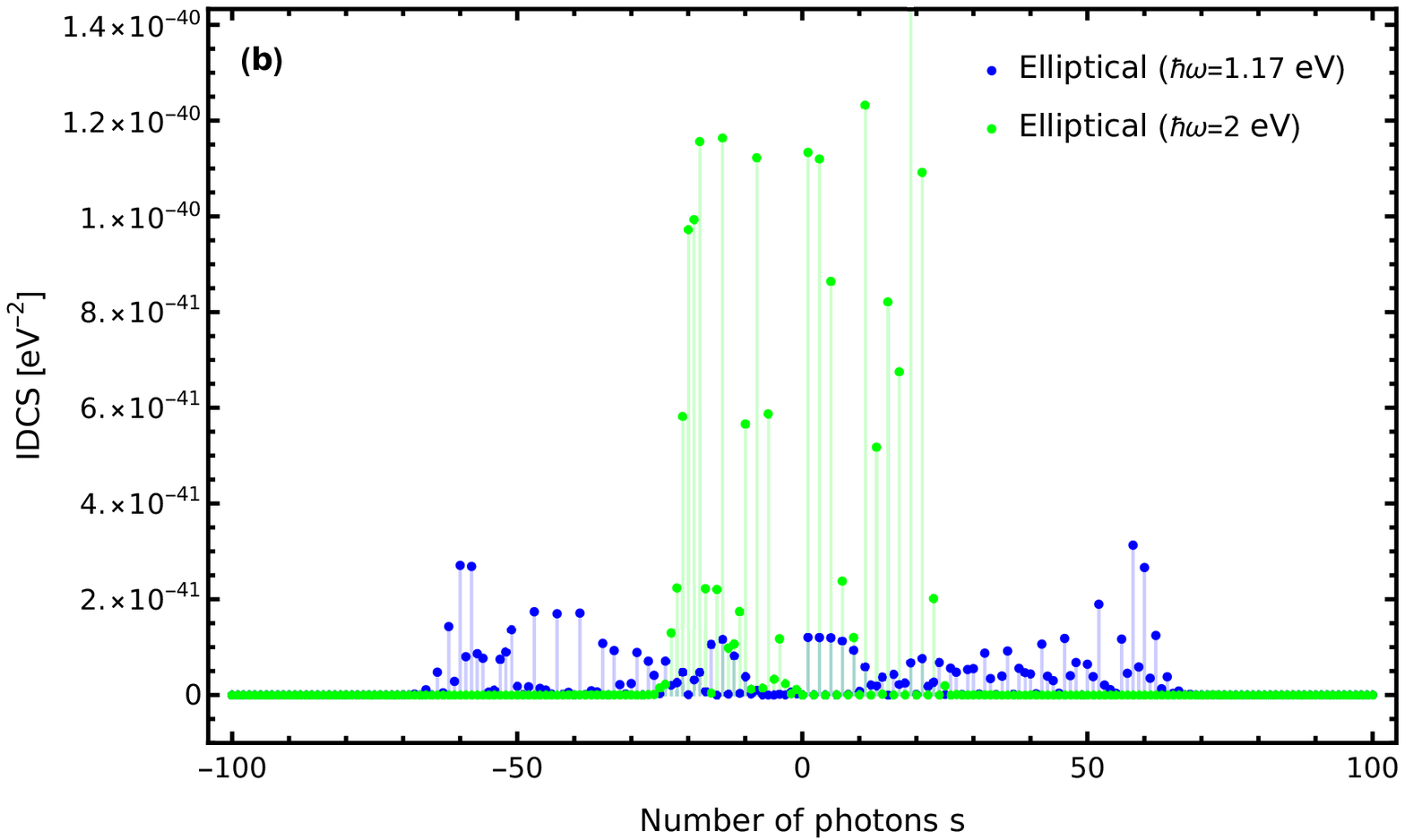} \par 
   \includegraphics[scale=0.45]{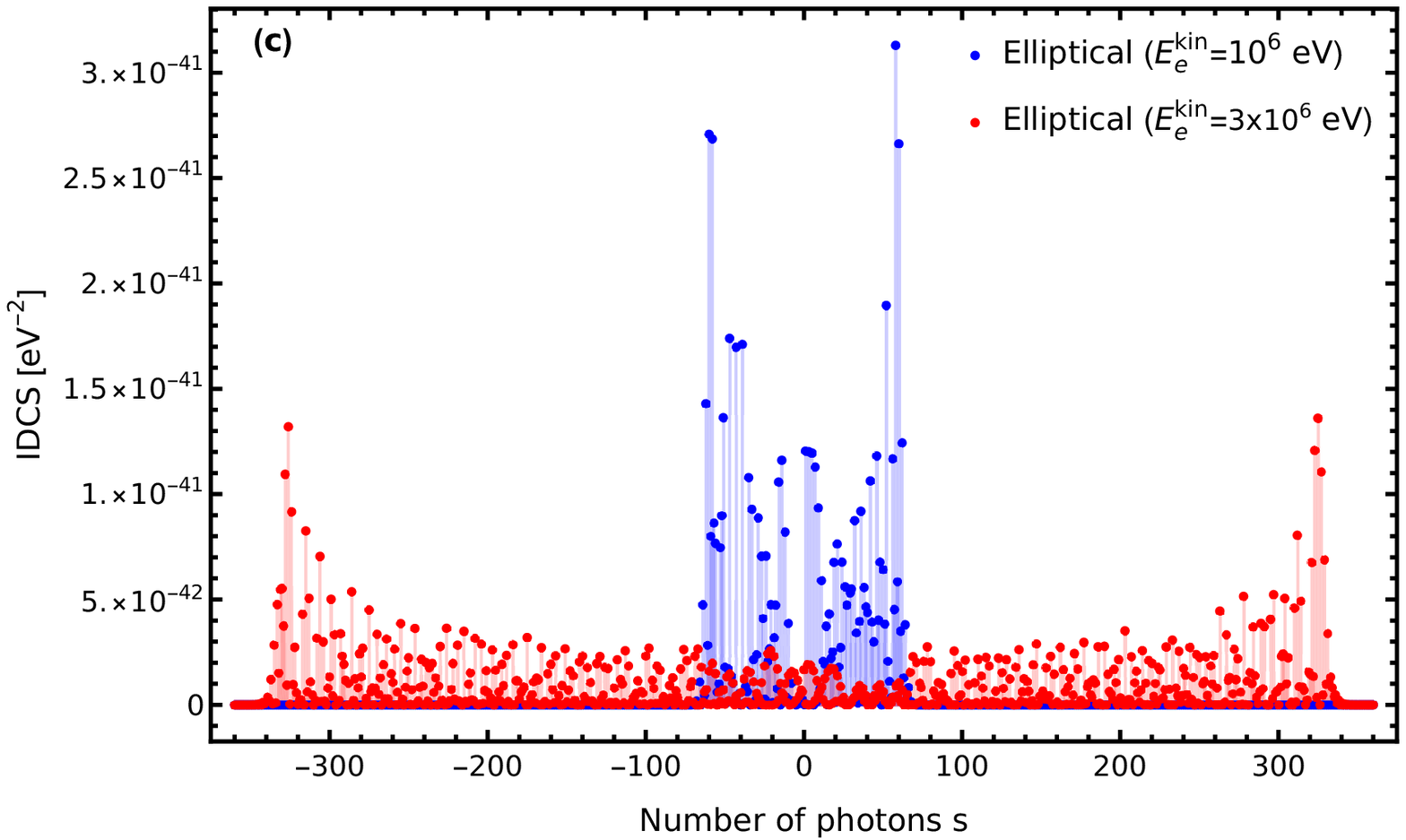}  
   \includegraphics[scale=0.5]{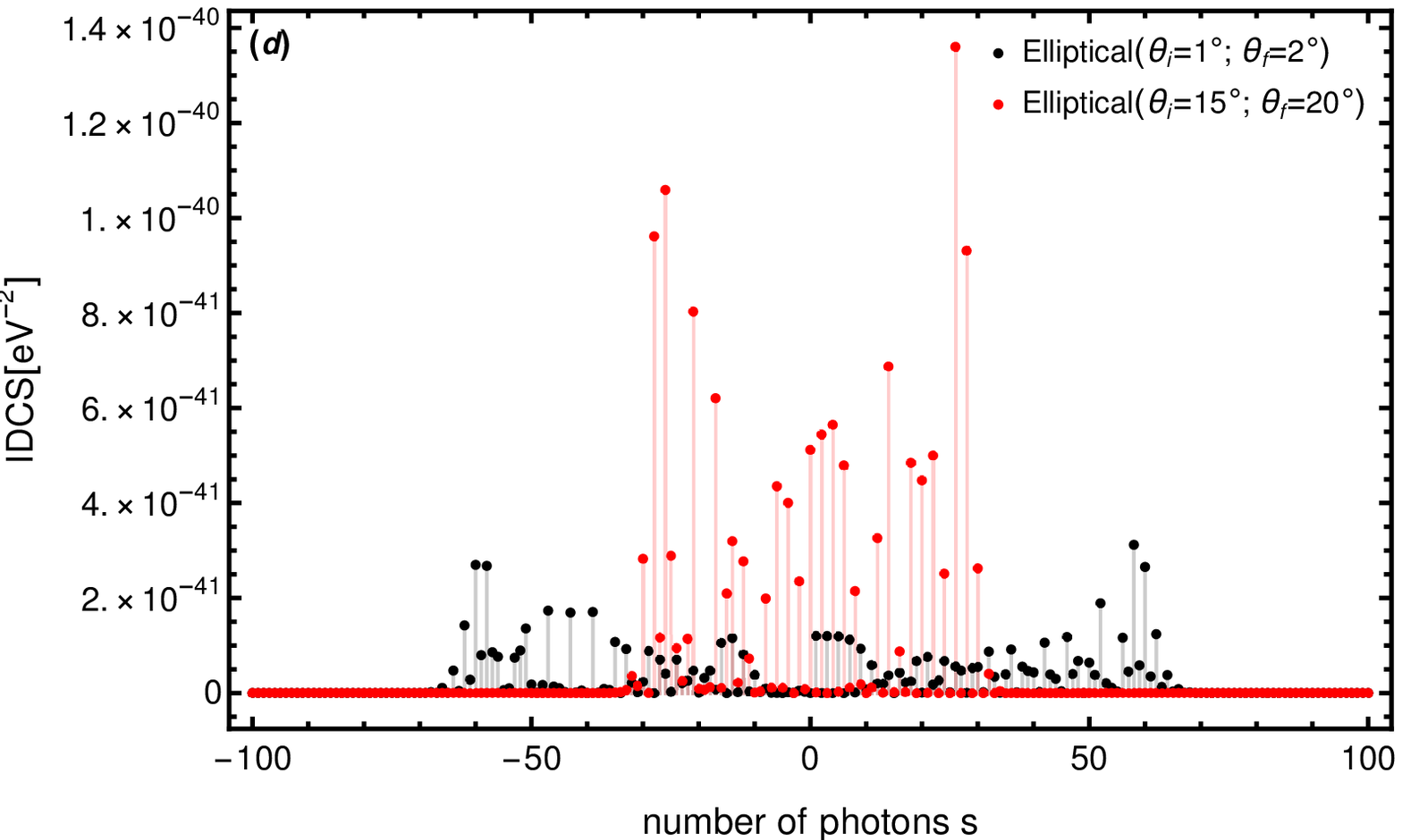}  
   \caption{The behavior of the IDCS, as a function of photon number $s$. The various parameters are (a) $\theta_{i}=1^{\circ}$, $\theta_{f}=2^{\circ}$ and $\hbar\omega=1.17$ eV, (b) $\mathcal{E}_{0}=5.18\times10^{8}$ V/cm, $\theta_{i}=1^{\circ}$ and $\theta_{f}=2^{\circ}$, (c)  $\mathcal{E}_{0}=5.18\times10^{8}$ V/cm, $\hbar\omega=1.17$ eV, $\theta_{i}=1^{\circ}$ and $\theta_{f}=2^{\circ}$ and (d) $\mathcal{E}_{0}=5.18\times10^{8}$ V/cm and $\hbar\omega=1.17$ eV.}\label{fig.IDCS-elliptic}
    \end{figure} 
In Fig.~\ref{fig.IDCS-elliptic}, we display the variations of the IDCS (the multi-photon energy transfer phenomenon) as a function of the number of photons exchanged $s$, at different field strengths and frequencies, and at different kinetic energies and geometry of the incident electron. From Fig.~\ref{fig.IDCS-elliptic}(a), we can observe that the electron exchanges a large number of photons with the high-intensity EM field ($\mathcal{E}_{0}= 10^{9}$ V/cm), where the cutoff number is approximately $s = \pm 125$, compared to the low-intensity EM field ($\mathcal{E}_{0}= 5.18 \times 10^{8}$ V/cm), where the cutoff number is approximately equal to $s= \pm 65$. This implies that the influence of the laser field on the scattering process is more significant at higher field strengths, i.e. the electron interacts powerfully with the strong EM field. Additionally, the order of magnitude of the IDCS decreases with the increase of the field strength. In Fig.~\ref{fig.IDCS-elliptic}(b), we visualize the variations of the IDCS as a function of the number of photons exchanged $s$ at different frequencies of the laser field. We see that at low frequencies ($ \hbar \omega= 1.17$ eV), the number of exchanged photons is large and the cutoff number is about $s = \pm 65 $, compared to high frequencies ($\hbar \omega= 2$ eV) where the cutoff number is about $s = \pm 25 $. Moreover, the order of magnitude of the IDCS increases with the increase of the frequency. Consequently, the multi-photon energy transfer phenomenon between the laser and the scattering process is related to the properties of the applied laser field. Figure~\ref{fig.IDCS-elliptic}(c) describes the variations of the IDCS as a function of the number of photons exchanged $s$ at different kinetic energies of the incident electron. The multi-photon energy transfer phenomenon increases as the kinetic energy of the incident electron increases, leading to an exchange of a large number of photons, while the order of magnitude of the IDCS decreases. In Fig.~\ref{fig.IDCS-elliptic}(d), we illustrate the influence of the chosen geometry on the multi-photon process. We observe that the number of photons exchanged in the geometry ($\theta_{i}=1^{\circ},~\theta_{f}=2^{\circ}$ ) is more important than the one exchanged in the geometry ($\theta_{i}=15^{\circ},~\theta_{f}=20^{\circ}$). While the first geometry means that the incident electron is almost in the same direction as the $\textbf{k}$ field vector, the second one indicates that the electron arrives at an angle of $\theta_{i}=15^{\circ}$ with respect to the $z$-axis. Physically, this states that the electron interacts with the laser if they arrive together in the same direction more than if they are in two different directions. We also see that the oscillations of these envelopes fall sharply to the sides, which can also be best explained by the well-known behavior and properties of ordinary and generalized Bessel functions (GBF)\cite{korsch2006two}.
Let us now see what happens if we sum over all the possible number of photons exchanged. In Fig.~\ref{fig.9-SDCS}, we have plotted the variations of the SDCS as a function of the final scattering angle $\theta_{f}$, for different known frequencies and electric field strengths.
\begin{figure}[hbtp]
 \centering
\includegraphics[scale=0.45]{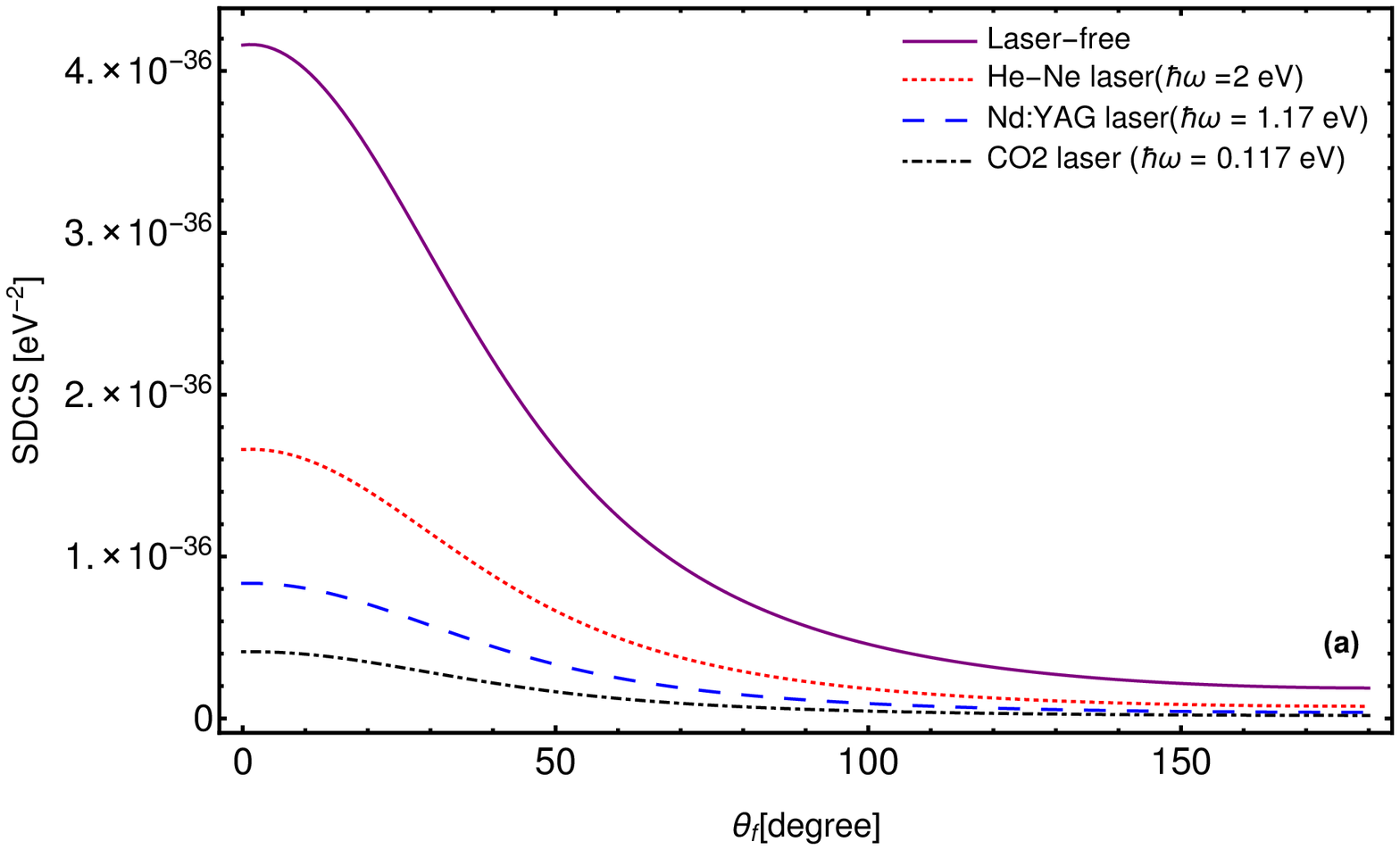}
\includegraphics[scale=0.45]{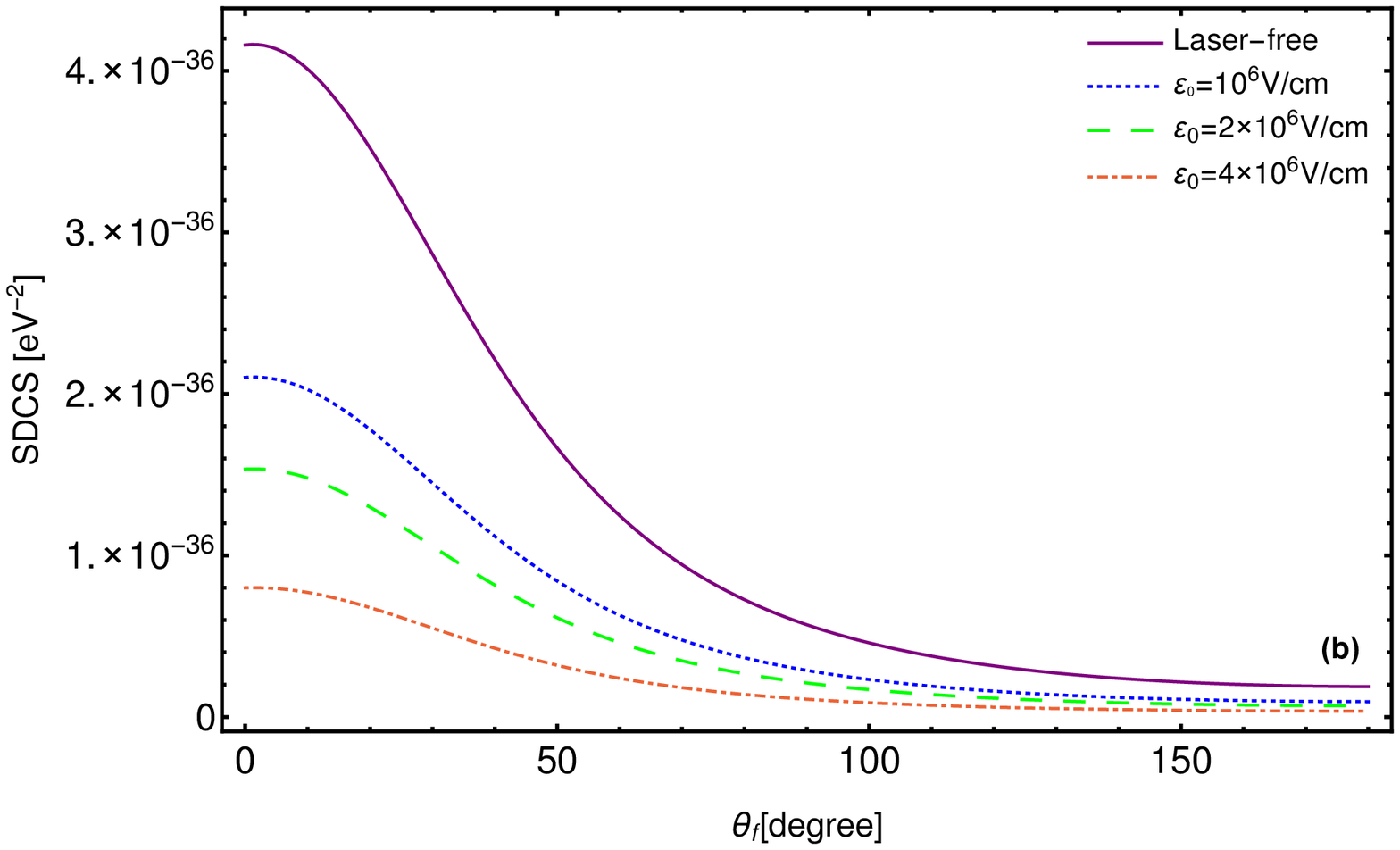}
\caption{Variation of the laser-assisted SDCS (with the degree of ellipticity $\eta=120^{\circ}$) as a function of the scattering angle $\theta_{f}$ for various electric field strengths $\mathcal{E}_{0}$ and laser frequencies $\hbar\omega$. The kinetic energy of the incident muon neutrino and the incoming electron are, respectively, $E_{\nu}^{\text{kin}}=0.5\times10^{-3}$ GeV et $E_{e}^{\text{kin}}=10^{-3}$ GeV. The other parameters are taken as follows: (a) $\mathcal{E}_{0}=10^{8}~\text{V/cm}$, $\theta_{i}=\varphi_{i}=\varphi_{f}=1^{\circ}$, (b) $\hbar\omega=1.17~\text{eV}$, $\theta_{i}=\varphi_{i}=\varphi_{f}=1^{\circ}$.}\label{fig.9-SDCS}
\end{figure} 
 Figure~\ref{fig.9-SDCS}(a) shows the dependence of the  SDCS assisted by elliptically polarized laser field with a degree of ellipticity $\eta=120^{\circ}$ as a function of the final scattering angle $\theta_{f}$, for different known laser frequencies $\hbar\omega$ which are the CO$_{2}$ laser ($\hbar\omega=0.117$ eV), the Nd:YAG laser ($\hbar\omega=1.17$ eV) and the He-Ne laser ($\hbar\omega=2$ eV). For $ 0^{\circ} \leqslant \theta_{f}\leqslant 180^{\circ}$, the order of magnitude of the SDCS increases with increasing laser field frequency, which confirms the result of Fig.~\ref{fig.IDCS-elliptic}(b). Since the laser field strength is also a crucial parameter, we performed the same analysis for different values of the laser field strength in Fig.~\ref{fig.9-SDCS}(b). We observe that the order of magnitude of SDCS decreases with increasing laser field strength, which is consistent with the result obtained in Fig.~\ref{fig.IDCS-elliptic}(a).

\begin{figure}[hbtp]
\centering
   \includegraphics[scale=0.45]{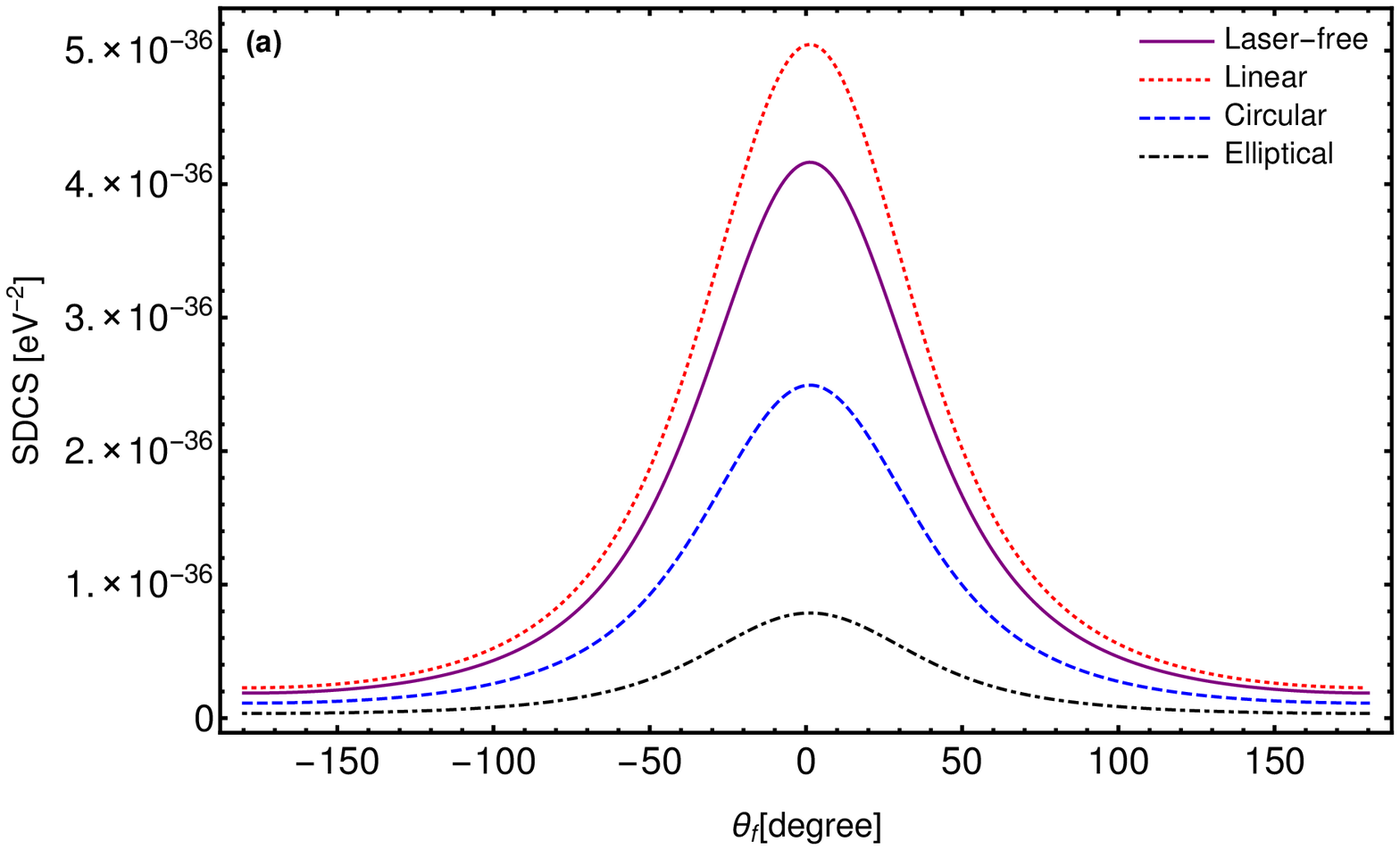}  
   \includegraphics[scale=0.45]{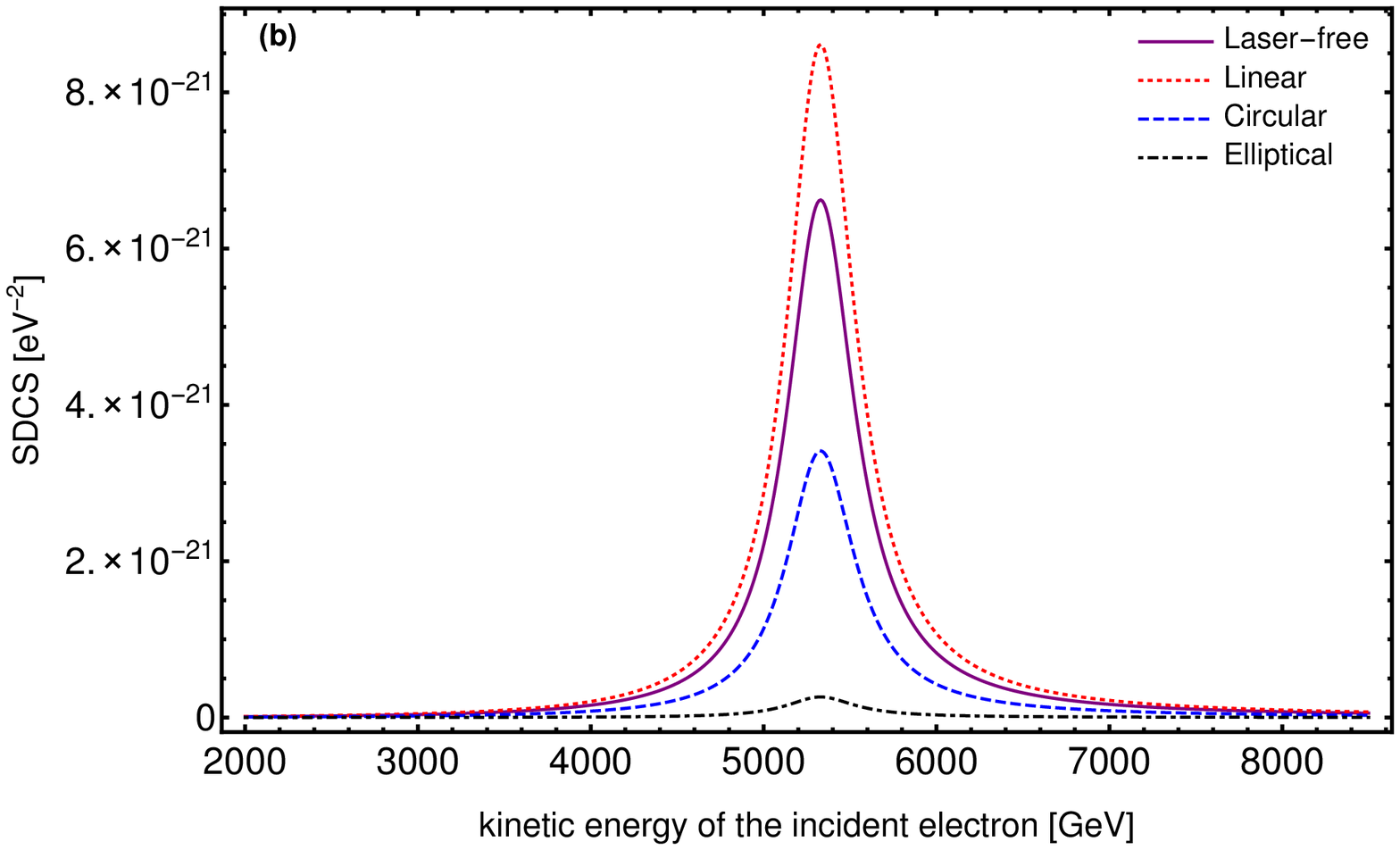}  
   \caption{Variations of the SDCS as a function of the scattering angle  $\theta_{f}$ and the kinetic energy of the incident electron for different polarizations of the strong EM field. The laser field parameters are selected as follows: $\hbar\omega=1.17~\text{eV}$ and $\mathcal{E}_{0}=10^{10}~\text{V/cm}$. The other parameters are taken as: (a) $E_{\nu}^{\text{kin}}=0.5\times10^{-3}~\text{GeV}$, $E_{e}^{\text{kin}}=10^{-3}~\text{GeV}$, $\theta_{i}=\varphi_{i}=\varphi_{f}=1^{\circ}$; (b) $E_{\nu}^{\text{kin}}=10~\text{GeV}$, $\theta_{i}=\varphi_{i}=\varphi_{f}=1^{\circ}$, $\theta_{f}=2^{\circ}$.}\label{fig.SDCS}
    \end{figure}

\begin{figure}[hbtp]
\centering
   \includegraphics[scale=0.53]{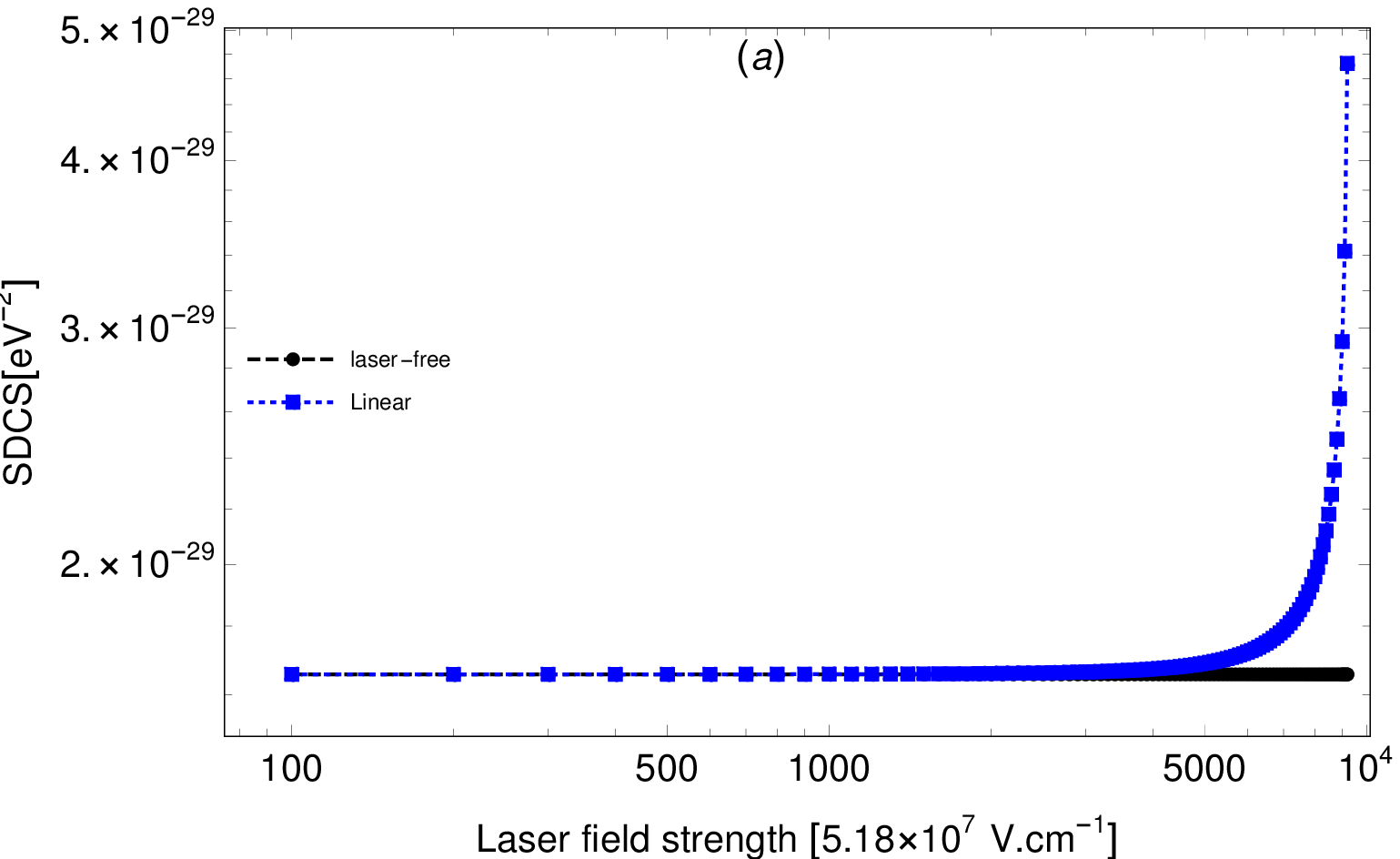}  
   \includegraphics[scale=0.55]{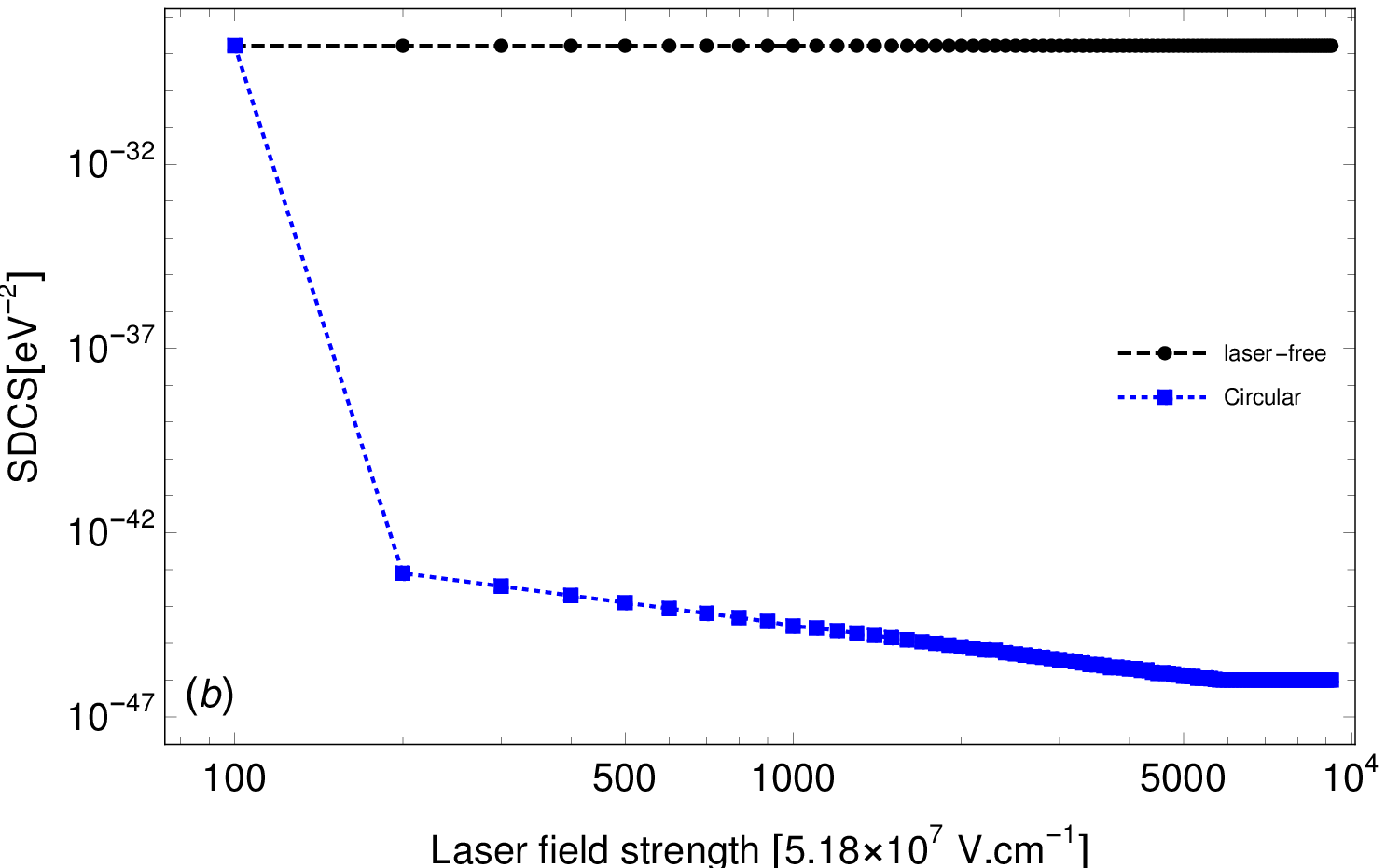}  \par
   \includegraphics[scale=0.55]{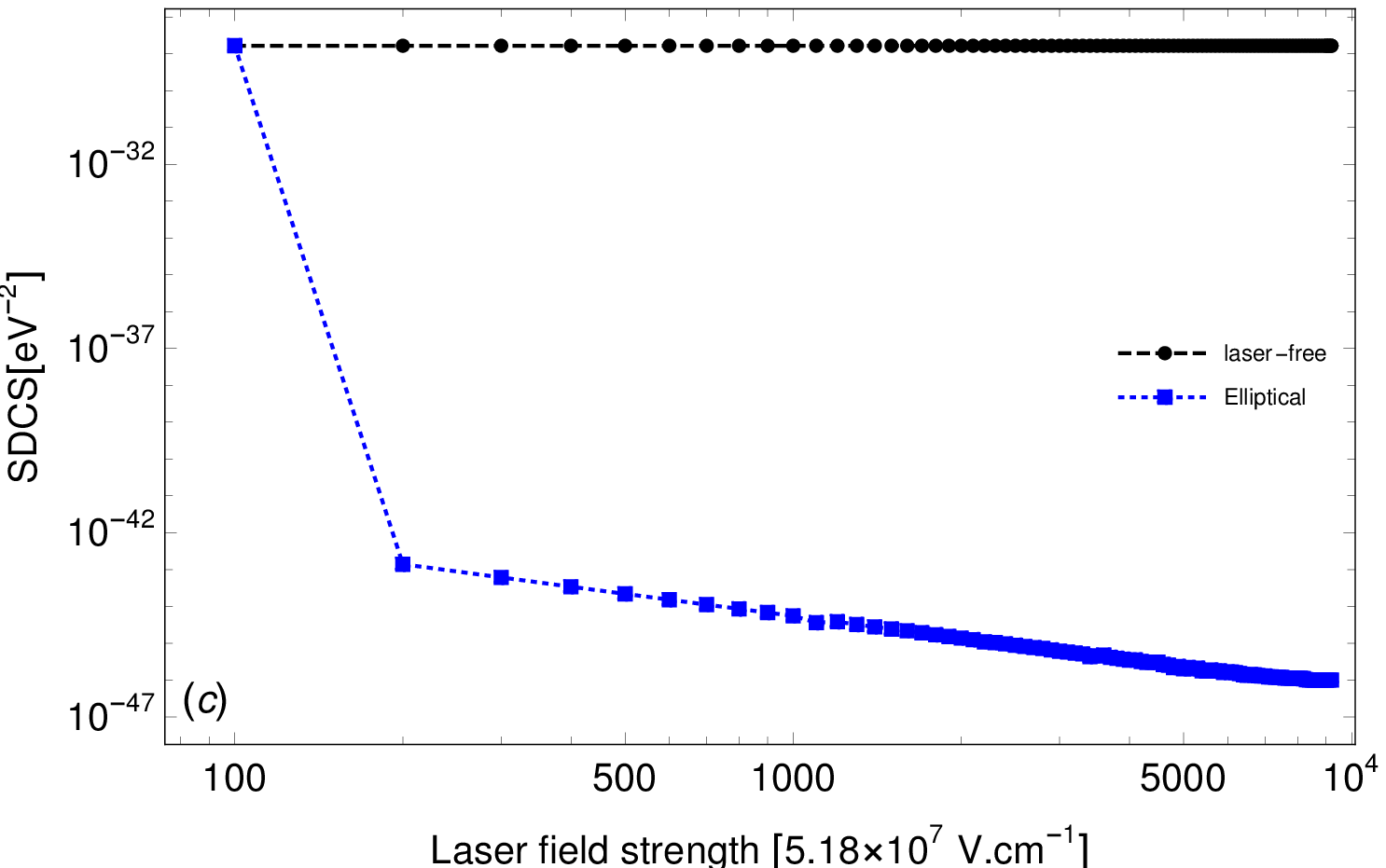}  
   \caption{Variations of SDCS as a function of the laser field strength $\mathcal{E}_{0}$ for different polarizations. The kinetic energy of the incident muon neutrino and the incoming electron are respectively $E_{\nu}^{\text{kin}}=0.5\times10^{-3}$ and $E_{e}^{\text{kin}}=10$ GeV. The frequency of the laser field is $\hbar \omega =0.117$ eV. The other parameters are chosen as follows: $\theta_{i}=\varphi_{i}=\varphi_{f}=1^{\circ}$ and $\theta_{f}=2^{\circ}$.}\label{fig.SDCS-intensitie}
    \end{figure}

Let us now turn to the discussion of the laser field polarization effect on the SDCS. Figure~\ref{fig.SDCS} shows the variations of the SDCS as a function of the final scattering angle $\theta_{f}$ and the kinetic energy of the incident electron $E_{e}^{\text{kin}}$ at different polarizations of the strong EM field. The parameters of the laser field are chosen as follows: $\hbar \omega=1.17~\text{eV}$ and $\mathcal{E}_{0}=10^{10}~\text{V/cm}$. Our results show that the effect of the laser field polarization is clearly highlighted, since the three SDCSs with laser and without laser are now obviously distinguishable. 
We also see that the SDCS in the case of linear polarization ($\eta=0^{\circ}$) is enhanced compared to that without laser. Furthermore, we observe that there is a reduction of the SDCS in the case of an elliptical ($\eta=120^{\circ}$) and circular ($\eta=90^{\circ}$) polarizations. We note here that the SDCS for elliptical polarization is lower or higher than the SDCS for circular polarization depending on the value of the degree of ellipticity $\eta$. The same thing can be said about the result presented in Fig.~\ref{fig.SDCS}(b) as a function of the kinetic energy of the incident electron $E_{e}^{\text{kin}}$.\\
Finally, to understand most clearly the effect of polarization, we display the dependence of the SDCS on the laser field strength at different polarizations in Fig.~\ref{fig.SDCS-intensitie}. The laser field strength appears in the equations for determining the behavior of the SDCS through the arguments $\xi_{1}$, $\xi_{2}$ and $\xi_{3}$ (see Eq.~(\ref{eq.11})) of the ordinary and generalized Bessel functions and the coefficients $\chi_{0 \mu}$, $\chi_{1 \mu}$, $\chi_{2 \mu}$ and $\chi_{3 \mu}$ given by Eq.~(\ref{eq.13}). 
In Fig.~\ref{fig.SDCS-intensitie}(a), we can see that the SDCS for linear polarization ($\eta=0^{\circ}$) remains constant and equal to the DCS without laser in the field strength range between $5.18\times10^{9}$ and $6. 21\times10^{10}$ V/cm. Outside this interval, we find that the SDCS increases progressively with the laser field strengh, which is consistent with the results obtained previously in \cite{bai2012multiphoton, du2018new, wang2019relativistic, Li-Berakdar2003}. Although the ordinary and generalized Bessel functions in the SDCS vary with the laser field strength, the electron state is more perturbed, and thus the cross section is strongly changed. Consequently, the SDCS increases because the energy of the electron increases after the absorption of laser photons. On the other hand, in Figs.~\ref{fig.SDCS-intensitie}(b) and \ref{fig.SDCS-intensitie}(c), which illustrate the variations of the  SDCS in the case of circular ($\eta=90^{\circ}$) and elliptical ($\eta=60^{\circ}$) polarizations, it is shown that the SDCS decreases and remains lower than the DCS without laser, which is consistent with the results found in \cite{attaourti2004mott, schnez2007laser}. Therefore, the effect of electron dressing induces very significant changes in the DCS, and a large enhancement is found in the linear polarization case.
\section{Conclusion} \label{CONCLUSION}
Laser-assisted electron-muon neutrino scattering is investigated, for elliptical polarization, in the first Born approximation and in the framework of electroweak theory.  We have extended the study of this scattering process for a general polarization that leads to all particular results obtained before in linear and circular polarizations. The numerical results show that the SDCS is significantly modified by the polarization type depending on the degree of ellipticity and the laser field parameters. Moreover, the SDCS in the case of linear polarization is enhanced compared to that without laser, while it is reduced in the case of elliptical and circular polarizations. We have shown that the scattering geometry, as well as the laser field parameters and the kinetic energy of the incident electron influence the multi-photon process. We hope that the present work will serve as a stimulus for experimenters to perform such scattering experiments on electrons dressed in linear and elliptical polarization.
\appendix*
\section{Explicit expression of the first four coefficients in Eq.~(\ref{trace-result})}\label{appendix}
The expressions for the coefficients $ \Lambda_{1} $ to $ \Lambda_{16}$ have been computed using the FeynCalc package. In order to limit the length of our article, we give here the first four coefficients $ \Lambda_{1} $ to $ \Lambda_{4} $: 
 \begin{equation}
\begin{split}
\Lambda_{1} =& 64 a^2 g_{A}^2 (k_{f}.k_{i}) \tan^2 (\eta/2) e^2 + (
   64 a^2 g_{A}g_{V} (k.k_{i}) (k_{f}.p_{f}) \tan^2 (\eta/2) e^2)/(k.p_{f}) \\ &
  - (32 a^2 g_{V}^2 (k.k_{i}) (k_{f}.p_{f}) \tan^2 (\eta/2) e^2)/(k.p_{f})- (
   32 a^2 g_{A}^2 (k.k_{i}) (k_{f}.p_{f}) \tan^2 (\eta/2) e^2)/(k.p_{f}) \\ &- (
   32 a^2 g_{A}^2 (k.k_{i}) (k_{f}.p_{i}) \tan^2 (\eta/2) e^2)/(k.p_{i}) - (
   64 a^2 g_{A}g_{V} (k.k_{i}) (k_{f}.p_{i}) \tan^2 (\eta/2) e^2)/(k.p_{i})\\ &- (
   32 a^2 g_{V}^2 (k.k_{i}) (k_{f}.p_{i}) \tan^2 (\eta/2) e^2)/(k.p_{i}) - (
   32 a^2 g_{A}^2 (k.k_{f}) (k_{i}.p_{f}) \tan^2 (\eta/2) e^2)/(k.p_{f})\\ & - (
   32 a^2 g_{V}^2 (k.k_{f}) (k_{i}.p_{f}) \tan^2 (\eta/2) e^2)/(k.p_{f}) + (
   64 a^2 g_{A}g_{V} (k.k_{f}) (k_{i}.p_{i}) \tan^2 (\eta/2) e^2)/(k.p_{i})\\ & - (
   32 a^2 g_{V}^2 (k.k_{f}) (k_{i}.p_{i}) \tan^2 (\eta/2) e^2)/(k.p_{i}) - (
   32 a^2 g_{A}^2 (k.k_{f}) (k_{i}.p_{i}) \tan^2 (\eta/2) e^2)/(k.p_{i}) \\ &- (
   64 a^2 g_{A}g_{V} (k.k_{f}) (k_{i}.p_{f}) \tan^2 (\eta/2) e^2)/(k.p_{f}) + (
   64 a^2 g_{A}^2 m^2 (k.k_{f}) (k.k_{i}) \tan^2 (\eta/ 2) e^2)/((k.p_{f}) (k.p_{i})) \\   
   & - (64 a^2 g_{V}^2 m^2 (k.k_{f}) (k.k_{i}) \tan^2 (\eta/2) e^2)/((k.p_{f}) (k.p_{i})) +
   (64 a^2 g_{A}^2 (k.k_{f}) (k.k_{i}) (p_{f}.p_{i}) \tan^2 (\eta/2) \\ &\times e^2)/((k.p_{f})    
   (k.p_i))  + 64 gA^2  m^2 (k_{f}.k_{i}) -  64 g_{V}^2 m^2 (k_{f}.k_{i}) + 64 g_{A}^2
   (k_f.p_i)(k_{i}.p_{f}) + 64 g_{V}^2 (k_{f}.p_{i}) (k_{i}.p_{f}) \\ & - 128 g_{A}g_{V}
   (k_{f}.p_{i})(k_{i}.p_{f}) + 64 g_{A}^2 (k_f.p_{f}) (k_{i}.p_{i}) + 64 g_{V}^2
   (k_{f}.p_{f}) (k_{i}.p_{i}) + 128 g_{A}g_{V} (k_{f}.p_{f}) (k_{i}.p_{i}) \\ & + 64 a^2
   g_{V}^2 (k_{f}.k_{i}) \tan^2 (\eta/2) e^2 +(64 a^2 g_{V}^2 (k.k_{f}) (k.k_{i})
   (p_{f}.p_{i}) \tan^2 (\eta/2) e^2) /((k.p_{f}) (k.p_{i})) \\ & +(32 a^4  g_{A}^2 (k.k_{f})
   (k.k_{i}) \tan^4 (\eta/2) e^4)/((k.p_{f}) (k.p_{i}))  + (32 a^4 g_{V}^2 (k.k_{f})
   (k.k_{i}) \tan^4 (\eta/2) e^4)/((k.p_{f}) (k.p_{i})),
  \end{split}
\end{equation}
\begin{equation}
\begin{split}
\Lambda_{2} = &(32 g_{V}^2 a^2 (k.k_f) (k_i.p_f) e^2)/(k.p_f)+(32 g_{A}^2 a^2 (k.k_{i})
 (k_f.p_f) e^2)/(k.p_f) + (32 g_{A}^2 a^2 (k.k_f) (k_i.p_f) e^2)/(k.p_f) \\ &+ 
 (32 g_{V}^2 a^2 (k.k_{i}) (k_f.p_f) e^2)/(k.p_f)-(64 g_{A}g_{V}a^2 (k.k_{i}) (k_f.p_f) e^2)/
 (k.p_f)-(32 g_{A}^2 a^2 (k.k_{i}) (k_f.p_f) e^2)/(k.p_i)\\ & 
 - (32 g_{V}^2 a^2 (k.k_{i}) (k_f.p_f) e^2)/(k.p_i) - (64 g_{A}g_{V}a^2 (k.k_{i}) (k_f.p_f)
 e^2)/(k.p_i) - (32 g_{A}^2 a^2 (k.k_{i}) (k_f.p_i) e^2)/(k.p_f) \\ &- 
 (32  g_{V}^2 a^2 (k.k_{i}) (k_f.p_i) e^2)/(k.p_f) + (64 g_{A}g_{V}a^2 (k.k_{i}) (k_f.p_i)
 e^2)/(k.p_f) - 64 gA^2 a^2 (k_f.k_i) e^2 \\ & 
 - 64 g_{V}^2 a^2 (k_f.k_i) e^2 + 128 g_{A}^2 (a_1.k_f) (a_1.k_i) e^2 + 128 g_{V}^2 (a_1.k_f) (a_1.k_i)
 e^2 + ( 32 g_{A}^2 a^2 (k.k_f) (k_i.p_i) e^2)/(k.p_i) \\ & 
 + (32 g_{V}^2 a^2 (k.k_f) (k_i.p_i) e^2)/(k.p_i)-(64 g_{A}g_{V}a^2 (k.k_f) (k_i.p_i) e^2)/(k.p_i) + (
 32g_{A}^2 a^2 (k.k_{i}) (k_f.p_i) e^2)/(k.p_i) \\ &+ 
 (32 g_{V}^2 a^2 (k.k_{i}) (k_f.p_i) e^2)/(k.p_i)+(64 g_{A}g_{V}a^2 (k.k_{i}) (k_f.p_i) e^2)/(k.p_i) 
 + (64 g_{A}g_{V}a^2 (k.k_f) (k_i.p_f) e^2)/(k.p_f) \\ &- 
 (32 g_{A}^2 a^2 (k.k_f) (k_i.p_f) e^2)/(k.p_i)-(32 g_{V}^2 a^2 (k.k_f) (k_i.p_f) e^2)/(k.p_i) + (
 64 g_{A}g_{V}a^2 (k.k_f) (k_i.p_f) e^2)/(k.p_i) \\ &
 -(32 g_{A}^2 a^2 (k.k_f) (k_i.p_i) e^2)/(k.p_f)-(32 g_{V}^2 a^2 (k.k_f) (k_i.p_i) e^2)/(k.p_f) - (
 64 g_{A}^2 a^2 (k.k_f) (k.k_{i})\\ & \times
 (p_f.p_i) e^2)/((k.p_f) (k.p_i))-(64 g_{V}^2 a^2 (k.k_f) (k.k_{i}) (p_f.p_i) e^2)/((k.p_f) (k.p_i))
 -(64 g_{A}^2 (a_1.k_i) (a_1.p_f) \\ & \times (k.k_f) e^2)/(k.p_f) 
 -(64 g_{V}^2 (a_1.k_i) (a_1.p_f) (k.k_f) e^2)/(k.p_f)-(128 g_{A}g_{V} (a_1.k_i) (a_1.p_f)  (k.k_f) e^2)
 /(k.p_f) \\ & 
 - (64 g_{A}^2(a_1.k_f) (a_1.p_f) (k.k_i) e^2)/(k.p_f)-(64 g_{V}^2 (a_1.k_f) (a_1.p_f) (k.k_i) e^2)/(k.p_f) 
 + (128 g_{A}g_{V} (a_1.k_f) \\ & \times (a_1.p_f) (k.k_i) e^2)/(k.p_f)-(64 g_{A}^2 (a_1.k_i) (a_1.p_i) (k.k_f)
 e^2)/(k.p_i)- (64 g_{V}^2 (a_1.k_i) (a_1.p_i) (k.k_f)  e^2)/(k.p_i) \\ &+ (128 g_{A}g_{V} (a_1.k_i)
 (a_1.p_i) (k.k_f) e^2)/(k.p_i)  - (64 g_{A}^2 (a_1.k_f) (a_1.p_i) (k.k_i) e^2)/(k.p_i)- (64 g_{V}^2
 (a_1.k_f) (a_1.p_i) \\ & \times(k.k_i) e^2)/(k.p_i) - (64 g_{A}^2 m^2 a^2 (k.k_f) (k.k_i) e^2)/((k.p_f)
 (k.p_i)) + (64 g_{V}^2 m^2 a^2 (k.k_f) (k.k_i) e^2)/((k.p_f) (k.p_i)) \\ &+ (128 g_{A}^2 (a_1.p_f)
 (a_1.p_i) (k.k_f) (k.k_{i}) e^2)/((k.p_f) (k.p_i)) + (128 g_{V}^2 (a_1.p_f) (a_1.p_i) (k.k_f) (k.k_i)
 \\ & \times e^2)/((k.p_f) (k.p_i))  - (128 g_{A}g_{V} (a_1.k_f) (a_1.p_i) (k.k_{i}) e^2)/(k.p_i)- ( 64 g_{A}
 g_{V} a^2 (k.k_f) (k_i.p_i) e^2)/(k.p_f),
  \end{split}
\end{equation}
\begin{equation}
\begin{split}
\Lambda_{3} = &(32 g_{A}^2 a^2 (k.k_f) (k_i.p_f) \tan^2 (\eta/2) e^2)/(k.p_f) + (
   32 g_{V}^2 a^2 (k.k_f) (k_i.p_f) \tan^2 (\eta/2) e^2)/(k.p_f) \\ &+ (
   32 g_{A}^2 a^2 (k.k_i) (k_f.p_i) \tan^2 (\eta/2) e^2)/(k.p_i) + (
   32 g_{V}^2 a^2 (k.k_i) (k_f.p_i) \tan^2 (\eta/2) e^2)/(k.p_i)\\ & + (
   64 g_{A}g_{V}a^2 (k.k_i) (k_f.p_i) \tan^2 (\eta/2) e^2)/(k.p_i) - 
   64 g_{A}^2 a^2 (k_f.k_i) \tan^2 (\eta/2) e^2 - 64 g_{V}^2 a^2 \\ & \times (k_f.k_i) \tan^2 (\eta/2) e^2 
   + (64 g_{A}g_{V}a^2 (k.k_f) (k_i.p_f) \tan^2 (\eta/2) e^2)/(k.p_f) - (
   32 g_{A}^2 a^2 (k.k_f) (k_i.p_f) \\ & \times \tan^2(\eta/2) e^2)/(k.p_i) - (
   32 g_{V}^2 a^2 (k.k_f) (k_i.p_f) \tan^2 (\eta/2) e^2)/(k.p_i) + (
   32 g_{A}^2 a^2 (k.k_f) (k_i.p_i)\\ & \times \tan^2 (\eta/2) e^2)/(k.p_i) + (
   32 g_{A}^2 a^2 (k.k_i) (k_f.p_f) \tan^2 (\eta/2) e^2)/(k.p_f) + (
   32 g_{V}^2 a^2 (k.k_i) (k_f.p_f) \\ & \times \tan^2 (\eta/2) e^2)/(k.p_f) - (
   64 g_{A}g_{V}a^2 (k.k_i) (k_f.p_f) \tan^2 (\eta/2) e^2)/(k.p_f)+ 
   128 g_{A}^2 (a_2.k_f) (a_2.k_i)  \\ & \times \tan^2 (\eta/2) e^2 + 
   128 g_{V}^2 (a_2.k_f) (a_2.k_i) \tan^2 (\eta/2) e^2 - (
   32 g_{A}^2 a^2 (k.k_i) (k_f.p_f) \tan^2 (\eta/2) e^2)/(k.p_i) \\ & - (
   32 g_{V}^2 a^2 (k.k_i) (k_f.p_f) \tan^2 (\eta/2) e^2)/(k.p_i) - (
   64 g_{A}g_{V}a^2 (k.k_i) (k_f.p_f) \tan^2 (\eta/2) e^2)/(k.p_i)  \\ & - (
   32 g_{A}^2 a^2 (k.k_i) (k_f.p_i) \tan^2 (\eta/2) e^2)/(k.p_f) - (
   32 g_{V}^2 a^2 (k.k_i) (k_f.p_i) \tan^2 (\eta/2) e^2)/(k.p_f) \\ & + (
   64 g_{A}g_{V}a^2 (k.k_i) (k_f.p_i) \tan^2 (\eta/2) e^2)/(k.p_f) + (
   32 g_{V}^2 a^2 (k.k_f) (k_i.p_i) \tan^2 (\eta/2) e^2)/(k.p_i) \\ & - (
   64 g_{A}g_{V}a^2 (k.k_f) (k_i.p_i) \tan^2 (\eta/2) e^2)/(k.p_i) + (
   64 g_{A}g_{V}a^2 (k.k_f) (k_i.p_f) \tan^2 (\eta/2) e^2)/(k.p_i)\\ & - (
   32 g_{A}^2 a^2 (k.k_f) (k_i.p_i) \tan^2 (\eta/2) e^2)/(k.p_f) - (
   32 g_{V}^2 a^2 (k.k_f) (k_i.p_i) \tan^2 (\eta/2) e^2)/(k.p_f) \\ &- (
   64 g_{A}^2 a^2 (k.k_f) (k.k_i) (p_f.p_i) \tan^2 (\eta/2) e^2)/((k.p_f) (k.p_i))
    - (64 g_{V}^2 a^2 (k.k_f) (k.k_i) (p_f.p_i) \\ &\times \tan^2 (\eta/2) e^2)/((k.p_f) (k.p_i))
     - (64 g_{A}^2 (a_2.k_i) (a_2.p_f) (k.k_f) \tan^2 (\eta/2) e^2)/(k.p_f) - (
   64 gV^2 (a_2.k_i) \\ & \times(a_2.p_f) (k.k_f)  \tan^2 (\eta/2) e^2)/(k.p_f) - (
   128 g_{A}g_{V} (a_2.k_i) (a_2.p_f) (k.k_f) \tan^2 (\eta/2) e^2)/(k.p_f) \\ &- (
   64 g_{A}^2 (a_2.k_f) (a_2.p_f) (k.k_i) \tan^2 (\eta/2) e^2)/(k.p_f) - (
   64 g_{V}^2 (a_2.k_f) (a_2.p_f) (k.k_i) \tan^2 (\eta/2) e^2)/(k.p_f) \\ &+ (
   128 g_{A}g_{V} (a_2.k_f) (a_2.p_f) (k.k_i) \tan^2 (\eta/2) e^2)/(k.p_f) - (
   64 g_{A}^2 (a_2.k_i) (a_2.p_i) (k.k_f) \tan^2 (\eta/2) e^2)/(k.p_i) \\ &- (
   64 g_{V}^2 (a_2.k_i) (a_2.p_i) (k.k_f) \tan^2 (\eta/2) e^2)/(k.p_i) + (
   128 g_{A}g_{V} (a_2.k_i) (a_2.p_i) (k.k_f) \tan^2 (\eta/2) e^2)/(k.p_i) \\ &- (
   64 g_{A}^2 (a_2.k_f) (a_2.p_i) (k.k_i) \tan^2 (\eta/2) e^2)/(k.p_i) - (
   64 g_{A}^2 m^2 a^2 (k.k_f) (k.k_i) \tan^2 (\eta/2) e^2)/((k.p_f) (k.p_i)) \\ &
   + (64 gV^2 m^2 a^2 (k.k_f) (k.k_i) \tan^2 (\eta/2) e^2)/((k.p_f) (k.p_i))
    + (128 g_{A}^2 (a_2.p_f) (a_2.p_i) (k.k_f) (k.k_i)\\ &\times \tan^2 (\eta/2) e^2)/((k.p_f)  (k.p_i)) 
    + (128 g_{V}^2 (a_2.p_f) (a_2.p_i) (k.k_f) (k.k_i) \tan^2 (\eta/2) e^2)/((k.p_f) (k.p_i))
    \\ & - ( 64 gV^2 (a_2.k_f) (a_2.p_i) (k.k_i) \tan^2 (\eta/2) e^2)/(k.p_i) - (
   128 g_{A}g_{V} (a_2.k_f) (a_2.p_i) (k.k_i) \tan^2 (\eta/2) e^2)/(k.p_i)\\ & - (
   64 g_{A}g_{V}a^2 (k.k_f) (k_i.p_i) \tan^2 (\eta/2) e^2)/(k.p_f),
  \end{split}
\end{equation}
\begin{equation}
\begin{split}
\Lambda_{4}= & -((32a^4 g_{A}^2 (k.k_{f})(k.k_{i})\tan^4(\eta/2)e^4)/((k.p_{f})(k.p_{i})))-(32a^4 g_{V}^2 (k.k_{f})(k.k_{i})\tan^4 (\eta/2)e^4)/((k.p_{f})(k.p_{i})) \\
& +(64a^4 g_{A}^2 (k.k_{f})(k.k_{i})\tan^2 (\eta/2)e^4)/((k.p_{f})(k.p_{i}))-(32a^4 g_{A}^2 (k.k_{f})(k.k_{i})e^4)/((k.p_{f})(k.p_{i}))\\
&-(32a^4 g_{V}^2 (k.k_{f})(k.k_{i})e^4)/((k.p_{f})(k.p_{i})) +(64a^4 g_{V}^2 (k.k_{f})(k.k_{i})\tan^2 (\eta/2)e^4)/((k.p_{f})(k.p_{i})).
 \end{split}
\end{equation}

\end{document}